# The Next Decade of Telecommunications Artificial Intelligence

Ye Ouyang, Ph.D.[1], Lilei Wang, Ph.D.[1], Aidong Yang, Ph.D.[1], Maulik Shah[2] ,
David Belanger, Ph.D.[3, 4], Tongqing Gao [5], Leping Wei[6], Yaqin Zhang, IEEE Fellow[7]

[1]AsiaInfo Technologies (China) Ltd., Beijing 100193, China;
[2]Verizon Communications Inc, New York NY 10036, USA;
[3]AT&T Inc, Dallas TX 75202, USA;
[4]Stevens Institute of Technology, Hoboken NJ 07030, USA;
[5]China Mobile Communications Group Co. Ltd., Beijing 100032, China;
[6]China Telecom Group Corporation, Beijing 100033, China;
[7]Tsinghua University, Beijing 100084, China

**Abstract:** It has been an exciting journey since the mobile communications and artificial intelligence (AI) were conceived 37 years and 64 years ago. While both fields evolved independently and profoundly changed communications and computing industries, the rapid convergence of 5$^{th}$ generation mobile communication technology (5G) and AI is beginning to significantly transform the core communication infrastructure, network management and vertical applications. The individual roadmaps of mobile communications and AI in the early stage were firstly outlined, with a concentration to review the era from 3$^{rd}$ generation mobile communication technology (3G) to 5G when AI and mobile communications started to converge. With regard to telecommunications AI, the progress of AI in the ecosystem of mobile communications was further introduced in detail, including network infrastructure, network operation and management, business operation and management, intelligent applications towards business support system (BSS) & operation support system (OSS) convergence, verticals and private networks etc. Then the classifications of AI in telecom ecosystems were summarized along with its evolution paths specified by various international telecommunications standardization organizations. Towards the next decade, the prospective roadmap of telecommunications AI was forecasted. In line with 3$^{rd}$ generation partnership project (3GPP) and international telecommunication union - radio communications sector (ITU-R) timeline of 5G & 6$^{th}$ generation mobile communication technology (6G), the network intelligence following 3GPP and open radio access network (O-RAN) routes, experience and intent driven network management and operation, network AI signaling system, intelligent middle office based BSS, intelligent customer experience management and policy control driven by BSS & OSS convergence, evolution from service level agreement (SLA) to experience level agreement (ELA), and intelligent private network for verticals were further explored. It concludes that with the vision AI will reshape the future B5G/6G landscape, and we need pivot our research and development (R&D), standardizations, and ecosystem to fully take the unprecedented opportunities.

# 1 MOBILE COMMUNICATION AND AI

The commercial development of mobile communication technology has gone through 37 years. It started from the commercialization of analog voice communication technology, advanced mobile phone system (AMPS) from Bell Laboratory and Motorola in Oct. 1983, to the 2G global system for mobile communications (GSM) which achieved the full digital voice telephony in 1991, and evolved to the 3G universal mobile telecommunications system (UMTS) in 2001 which supported the mobile Internet solutions, and then to the 4G long term evolution (LTE) with global large-scale commercialization so far which supports the all-internet protocol (all-IP) broadband connectivity, and eventually developed into the 5G technology in 2018 and gradually became commercially available all over the world [1]. In the development of mobile communications over the past 30 years, it has evolved from analog to digital, from voice only to voice and data service, from circuit switching to all-IP, from enclosed communication ecosystem to open ecosystem empowering vertical industry. At the early stage of mobile communication development, especially in the initial phase from 1G to 3G, mobile communication network and service ecosystem still conducted continuous integrity construction. Till the 4G ecosystem basically achieves the all IP-based network system, supports the voice and data services and starts trying to empower the vertical industry, the telecom industry starts proposing the requirement of automation and intelligentization development of mobile communication network. As mobile communication network becomes more complex and communication services become more diverse, communication network infrastructure and service system need to confront many complex scenarios, including very complicated wireless environment which can't be simulated with data model, exponential increasing complexity of IP switching and route selection, active network support and service guarantee, customized network service of "one customer with one strategy" and "one moment with one strategy", etc. All of them far outperform the processing and management system, which is predefined and executed by traditional artificial rules. Therefore, current communication system needs a set of automatic and intelligent system and method to guarantee the operation and development of network and service.

In the past 20 years from the commercial use of 3G in 2001 to the extensive commercial use of 5G in 2020, the mobile internet and data service have been prosperous, and the massive big data generated in communication ecosystem provide natural and high-quality data source for the development of AI in communication field. While in 2006, Hinton et al. [2] proposed deep learning, which marked the emergence of the third wave for AI development. The supervised learning, unsupervised learning and reinforcement learning in traditional machine learning, and the neural network in AI are applied with deep learning in various scenarios of communication field. After the search of academic achievements in telecommunications AI field through IEEE Explore database, it is found that the quantity of academic papers searched through keywords such as "Artificial Intelligence", "Machine Learning", "Deep Learning", etc. increases 6.42 times from 2006 to present. Thus it can be seen that, since the third wave development of AI in 2006, the integrated application of AI and mobile communication industry has entered a fast-growing stage.

General purpose technology (GPT) usually refers to those technologies which can influence the global or national economy. GPT is promising to significantly change the society by influencing the existing economy and social structure [3]-[4]. Economists Richard Lipsey and Kenneth Carlaw defined 24 kinds of technologies, including AI, etc. as the GPT as early as in 2005[6]. Since 2018, governments and academic organizations around the world have gradually deemed 5G as a new generation of GPT [6]-[10]. GPT has technology diffusivity and empowerment in various industries, and can boost the productivity for the R&D and innovation of vertical industry [4]. 5G and AI own those characteristics very obviously. Therefore, 5G and AI are commonly regarded as a group of the latest GPT in the 21st century by various countries and adopted by industrial field.

Since 2018, the 5G has gradually been used for commercial purpose, and many papers [1][11][12] have explored the application of AI in 5G in the survey or demonstration study. Most of them summarize AI through the application cases of 5G physical layer, medium access control (MAC) layer, network layer and application layer, or conduct simulation or data analysis of certain research problem for field experiment or demonstration research. However, the overall review and forward-looking, which are rare in the industry currently, are needed for the integrated development of mobile communication and AI technology from the perspective of 5G and beyond 5G (B5G) ecosystem. The authors hope to deem the integrated application development of 5G and AI technology (a group of GPT) as main line, conducts the systematic review of "intelligence injection" and "intelligence integration" of AI in current 5G technology international standards, 5G network and service ecosystem, and conducts the forward-looking discussion of mobile communication and AI technology in the next decade.

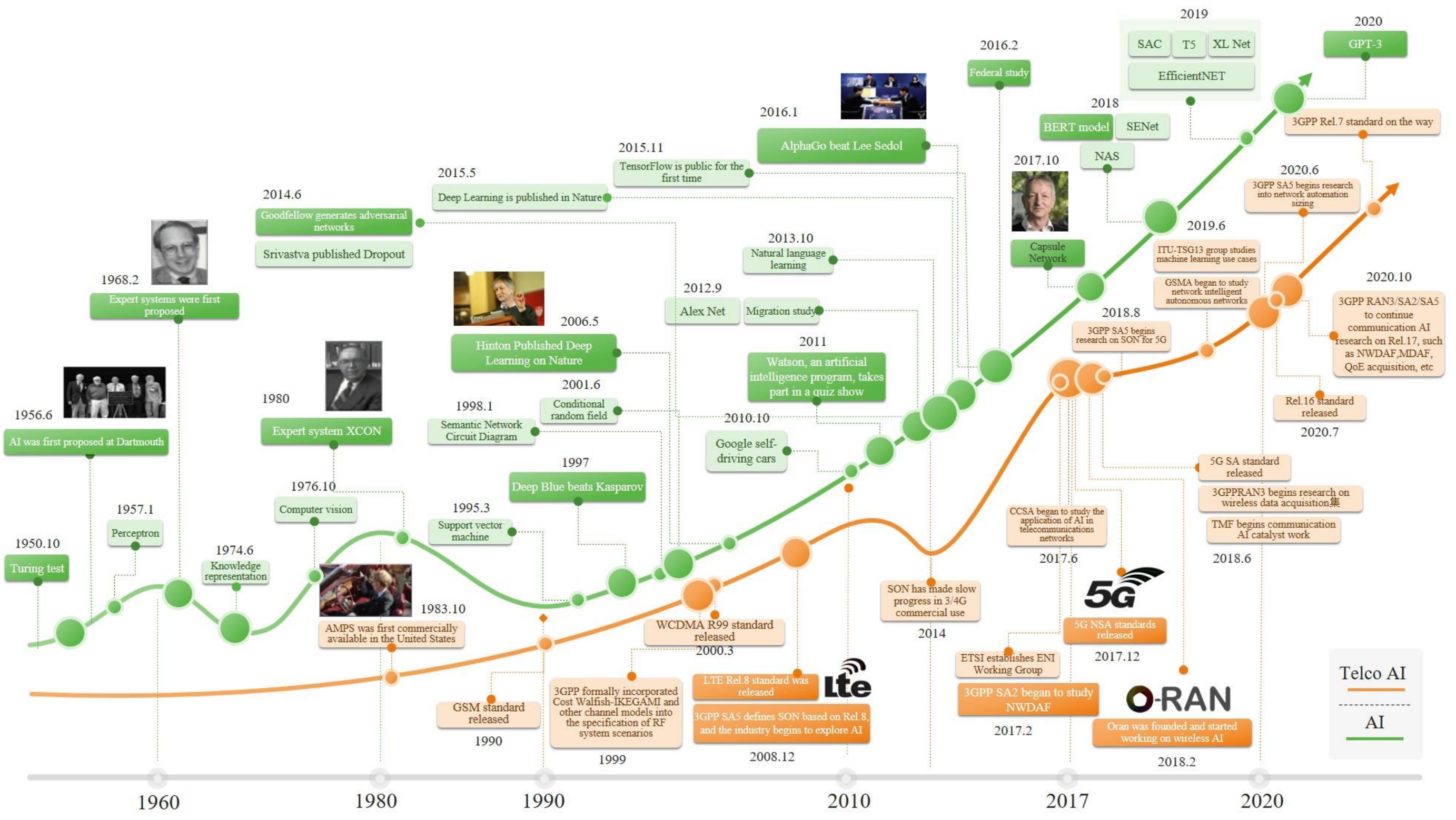

Figure 1 Comparison of Evolution Path between Communication Technology and AI Technology

# 2 DEVELOPMENT PATH OF MOBILE COMMUNICATION AND AI

Mobile communication technology and AI have distinct and independent development roadmap at their respective early stages. In the development of Mobile Communication technology from 2G to 5G, the industrial field basically evolves along the main line with 3GPP as fact technology standard and is supplemented by other technology standards such as European telecommunications standards institute (ETSI), ITU, open RAN alliance (O-RAN), etc. as the sideline. Since 2008, 3GPP has gradually introduced the AI concept to the technology standards of the mobile communication network, with self-organizing networks (SON) technology as a significant mark.

## 2.1 Historical Development Path of AI

AI technology occurred in 1956 at the earliest, and the word "AI" was raised at the United States Dartmouth Conference [13], and Arthur Samuel proposed the Computational Learning Theory in the same year, as shown in Figure 1. In the mid-1997s, bionics-based school of thought became popular, and the expert system, neural network, etc. obtained high-speed development [14]. Since then, people have attempted to research the AI program with universality, but have confronted severe obstacles, been caught in stagnation and entered the "severe winter" period of development [15]. In 1997, the success of "Deep Blue" contributed to the development planning of AI. With the increase of computing power and the massive data brought by internet popularization, the bottleneck of AI has been broken, providing development possibility for the deep learning and reinforcement learning based on big data. In the beginning of the 21st century, the AI technology develops from "perception" to "cognition", and obtains important progress in the deep learning technology such as voice processing, text analysis, video processing, etc. In 2012, Hinton released a convolutional network AlexNet with elaborate design [16], which added ReLU and Dropout processing method to traditional convolutional network and expanded network structure to larger scale, greatly reducing the error rate of image recognition. Natural language processing (NLP) obtained significant progress in 2013: Hinton group recognized voice with recurrent neural network (RNN), and Turing Award winner Yoshua Bengio proposed a kind of language model word2vec based on neural network in the same year for text analysis. Both techniques provide significantly better recognition results than traditional methods. Generative adversarial networks (GAN) technology born in 2014 is particularly concerned by the academic and industrial fields [17], the realistic effect of the latest GAN algorithm in image generation field has been unable to be discerned with human eyes. The deep reinforcement learning model DQN (Deep Q-Network) was published on Nature in 2015, which marked the milestones of reinforcement learning and deep learning [18]. In 2016, AlphaGo, which combines deep neural networks, reinforcement learning, and Monte Carlo tree search, was developed by Google DeepMind and defeated many Go champions. At the end of 2018, Google released a bidirectional language model Bidirectional Encoder Representations From Transformers (BERT) [19], which opened the "Pandora's box" of deep learning in NLP application and caused great interest in the industry, becoming an important stage for NLP technology development. In 2020, Open-AI developed the pre-trained model GPT-3 with 175 billion parameters based on the generative pre-trained transformer system, became the powerful general-purpose language model in the NLP field at present, and showed the capacity close to the mankind in the application tasks such as translation, Q&A, text gap filling, etc. [20]. In recent five years, the data privacy security has gradually attracted the attention of the world [21], and the "data silo" effect has become the "stumbling block" impeding the big data integration and AI development. In order to reconstruct industry data ecology, Federated Learning technology was proposed by Google for the first time at the end of 2017 [22], which broke the "data silo" deadlock by virtue of a kind of distributed encryption machine learning method. In 2018, in order to meet the industrial data joint requirements, WeBank proposed a kind of industrial-level Federated Learning framework -- FATE, which achieved a new paradigm of industrial Federated Learning. At the end of 2020, international standard IEEE P3652.1 about industrial Federated Learning was released, which marked the industrial "ecological alliance of intelligence fusion" was officially implemented.

## 2.2 Historical Development Path of Telecommunications AI

Influenced by algorithm, computing power, requirement, etc., early mobile communication systems (such as AMPS, GSM, etc.) didn't cover AI application. But the analytical method based on data model and simulation has been applied in network planning and optimization. In 1968, Yoshihisa Okumura proposed the Okumura model which conducted simulations of real wireless channel based on measured data, which can be deemed as the embryonic form that the mobile communication system applied the data science algorithm at the early stage [23]. In 1980, Masaharu Hata proposed the Hata model to optimize the Okumura model [24]. As shown in Figure 1, 3GPP officially included the COST Walfish-Ikegami and other channel models into the scenario specifications of 3G radio frequency systems in 1999 [25]. Later, with the development of wireless cellular technology, more wireless channel simulation models or algorithms occurred [26]-[28].

3GPP started defining SON function in 2008 [29][30], then communication community started exploring the application of various AI algorithms in SON. At the initial stage, the genetic algorithm, evolutionary algorithm, multi-objective optimization algorithm and other distributed optimization algorithms are mainly applied to optimize the network coverage and capacity [31][32]. Machine learning has been widely accepted by the SON field and is used as a key technology to achieve network's self-organization, self-configuration, self-optimization and autonomy [33]. However, truly rapid development of telecommunications AI started in 2017.

In Feb. 2017, 3GPP service & system aspects working group 2 (SA 2) started researching intelligent network functions of 5G core network: network data analytics function (NWDAF) [34], including UE (user equipment) mobile management, such as paging enhancement and connection management enhancement based on UE mobility model prediction, etc.; 5G quality of service (QoS) enhancement, such as parameter configuration optimization; network load optimization, such as user plane function (UPF) based on network performance prediction, etc. In the same month, ETSI established the experiential networked intelligence (ENI) working group which specialized in experiential perception network management architecture, case, terms, etc. [35]. In Jun. 2017, China Communications Standards Association (CCSA) started the research of Applied Project of AI in telecommunication network [36]. In Feb. 2018, Open Radio Access Network Alliance (O-RAN Alliance) was established and started formulating the framework, use cases, process and interface specifications of wireless AI [37]. In Jun. 2018, 3GPP Radio Access Network Working Group 3 (RAN3) started researching the data acquisition mechanism on wireless side [38]. Telecom Management Forum (TMForum) also began the catalyst work related to AI. In Aug. 2018, 3GPP SA5 started the topic research related to 5G SON [39]. In Oct. 2018, 3GPP SA5 began the research on AI and defined a new management plane function: management data analytic function (MDAF) [40]. In Jun. 2019, ITU Telecommunication Standardization Sector Study Group 13 (ITU-T SG13) started the research on machine learning use cases [41]. In the same month, Global System for Mobile Communications Association (GSMA) began the whitepaper work of intelligent autonomous network cases [42]. In Jun. 2020, 3GPP SA5 started the research topic about network automation classification [43]. In the same month, China Mobile united AsiaInfo Technologies to officially introduce the Federated Learning concept to 3GPP R17 standard for the first time, forming the first global international standard of Federated Learning in 5G field [44]. In Jul. 2020, 3GPP RAN3, SA2 and SA5 would continuously boost the research on AI-related standardized subjects such as NWDAF, MDAF, Quality of Experience (QoE), etc. after the 3GPP R16 was officially frozen.

# 3 DEVELOPMENT OF TELECOMMUNICATIONS AI

The essence of communication is to pass the user information to destination from starting point through various communication technologies (such as Mobile Communication, Satellite Communication, Fixed Network Communication, etc.). The quality of communication is measured by whether the information can be reproduced accurately and perfectly from the transmitter end to the receiver end. Compared with the natural intelligence of mankind and animals, AI makes computer or machine simulate human thinking and cognition, such as the capacity of "learning" and "problem solving", perceive environment and take the corresponding actions to smoothly achieve the presupposed objectives with the maximum probability [45][46]. Communication and signal processing system are constructed based on precise mathematical model. AI absorbs knowledge from data and makes decision through the deep learning in AI, without explicit mathematical modeling and analysis. Communication system easily breaks away from the actual condition in practical application in case of excessively exquisite and elegant mathematical assumptions. In turn, AI or deep learning, if applied to communication systems, has an excessively black-box learning process, it easily causes the communication and information model construction lacks physical meaning. The ranked autonomy is deemed as a typical feature of communication system and the inter-connection & inter-communication through standardized interactive interface is achieved to form complete system. The transmitter and receiver of signal processing system can be decomposed into different processing units which are responsible for their respective functions, such as information encoding and decoding, channel encoding and decoding, signal modulation and demodulation, etc., very similar with the micro-service concept in current IT system. Although such system architecture isn't the global optimal one, it independently analyzes and optimizes each subsystem to form an overall stable system. Modern Mobile Communication system has developed for over 30 years, so its efficiency and performance has been very excellent and has been close to Shannon Limit. Different from traditional layered autonomous method, the development of intelligent communication system may be pushed to a new stage if communication system is considered as a whole model through AI and deep learning for analysis and optimization. The development and application of AI in each communication ecosystem, as well as the staging system of current telecommunications AI development defined by communication international standards organizations.

## 3.1 Development of Telecommunications AI in Mobile Network Infrastructure

The development of AI in communication network infrastructure is expounded in four aspects: Radio Access Network, Core Network, Transport Network, and Terminal.

- Radio Access Network

The physical carrier of radio access network is base station. 5G base station is divided into the central unit (CU) and distributed unit (DU), which are similar as traditional baseband unit (BBU), connects active antenna unit (AAU) through optical fiber. AAU contains traditional Remote Radio Unit (RRU) and antenna function, namely the active RF part and the passive antenna. AI is applied in the physical layer, MAC layer and network layer through intelligence functions oriented to CU, DU and AAU in radio access network [47].

In respect of the physical layer and data link layer, the typical AI application includes the evaluation and prediction of the function of channel quality, orthogonal frequency division multiplexing (OFDM) symbolling, received signal detection, channel encoding and decoding, random access of dynamic frequency spectrum, etc. For example, the OFDM pilot signal analysis through DNN or reinforcement learning (RL) algorithm can help massive MIMO system obtain the complete and accurate channel state information (CSI). At receiver, the detection of OFDM symbol often depends on the maximum likelihood estimation, which is pretty sensitive to the CSI error and model accuracy, the papers [48][49] attempt the DNN algorithm, and the results indicate it surpasses the traditional method of MIMO symbol detection. In the 5G channel encoding and decoding, Polar code is used for channel control information and low-density parity-check (LDPC) is used for channel data transmission. Polar code needs to take multiple iterative computing for the optimal performance convergence, and the LDPC code has high complexity in decoding the large block codes with strong channel noise. Thus various decoding algorithms based on CNN, DNN and reinforcement learning [50]-[53] present outstanding performance and quality with low compute cost. Random access of dynamic frequency spectrum can also attempt to achieve the dynamic spectrum access of large-scale terminal based on the reinforcement learning based DSA strategy (learning-based random access and dynamic frequency spectrum access) in the future.

For the application layer of radio access network, 3GPP defines the standard system [54]-[64], in order to achieve the self-configuration, self-optimization and self-healing of wireless network. Although 3GPP doesn't specify or suggest any statistics or data model, suggests a series of intellectualized SON application scenarios: network coverage and capacity optimization (CCO), energy saving management (ESM), remote control of electrical tilting antennas (RET), interference reduction (IR), automated configuration of physical cell identity (ACPCI), mobility robustness/handover optimization (MRO/MHO), mobility load balancing (MLB), RACH optimization, automatic neighbor relation (ANR), inter-cell interference coordination (ICIC ), random access channel optimization (RACO), load-balancing optimization, self-healing functions, cell outage detection & compensation, minimization of drive-tests (MDT), etc. From 2008 when the first 3GPP SON was launched to present, data scientists in the industry [65]-[85] have attempted various AI models in order to achieve the SON application scenarios as described above. SON has plain development in 12 years from the 3G era to 5G era. Communication standard formulators hope to inject the automatic and intelligent gene into the mobile communication network architecture through SON regardless of distributed-SON (D-SON), centralized-SON (C-SON) and hybrid-SON (H-SON). Traditional communication equipment suppliers hope SON becomes a part of own network infrastructure for tight coupling, therefore, most of their SONs support own DE wireless network equipment. Telecom operators hope their radio access network is equipped with a set of intelligent SON system with neutral manufacturer and technology. The emerging SON startup company is willing to jointly devote to developing the SON technology with neutral network together with telecom operators, but suffers from the closure and non-standard of wireless equipment manufacturer in function interface and dada interface setting and can't achieve the 3GPP SON vision in the commercialization process. The relatively symbolic event in industrial field was that in Feb. 2013, the American Cisco spent USD 475 million in acquisition of Israeli Intucell which was a telecommunication software star company focusing on SON at that time. After the gloomy development for 7 years, Cisco sold relevant department to Indian HCL only at 10% of acquisition price (namely USD 50 million) in Jun. 2020. Telecom software company Amdocs ranking first in the world purchased the American SON Startup Company Celcite with USD 129 million in Nov. 2013, but later, the telecom operator Verizon and AT&T ranking first and second in the United States, respectively, failed to obtain prospective commercial market share in the commercial use of SON. The American mobile communication operator has also gradually adopted self-developed mode for attempt in the intelligent field of radio access network in recent 5 years, for example, Verizon's SON system jointly arranged with Cisco and Ericsson was gradually replaced by the own V-SON system in 2015.

- Core Network

The development of AI in the mobile communication core network obtains significant development in the 5G era. 3GPP SA2 defined the AI network function NWDAF in Feb. 2017, which was deemed to define, standardize and require the deployment of network AI in mobile communication core network architecture for the first time from 1G to 5G. The NWDAF framework is shown in Figure 2(a), the network function aims to integrate communication protocol by utilizing AI algorithms for the intelligent management and optimization of 5G core network mobile management, network QoS and the other network functions (such as UPF, etc.) and for the improvement of network QoS and QoE. Currently, the Chinese and American operators are conducting the function test of NWDAF for commercial use in 5G SA.

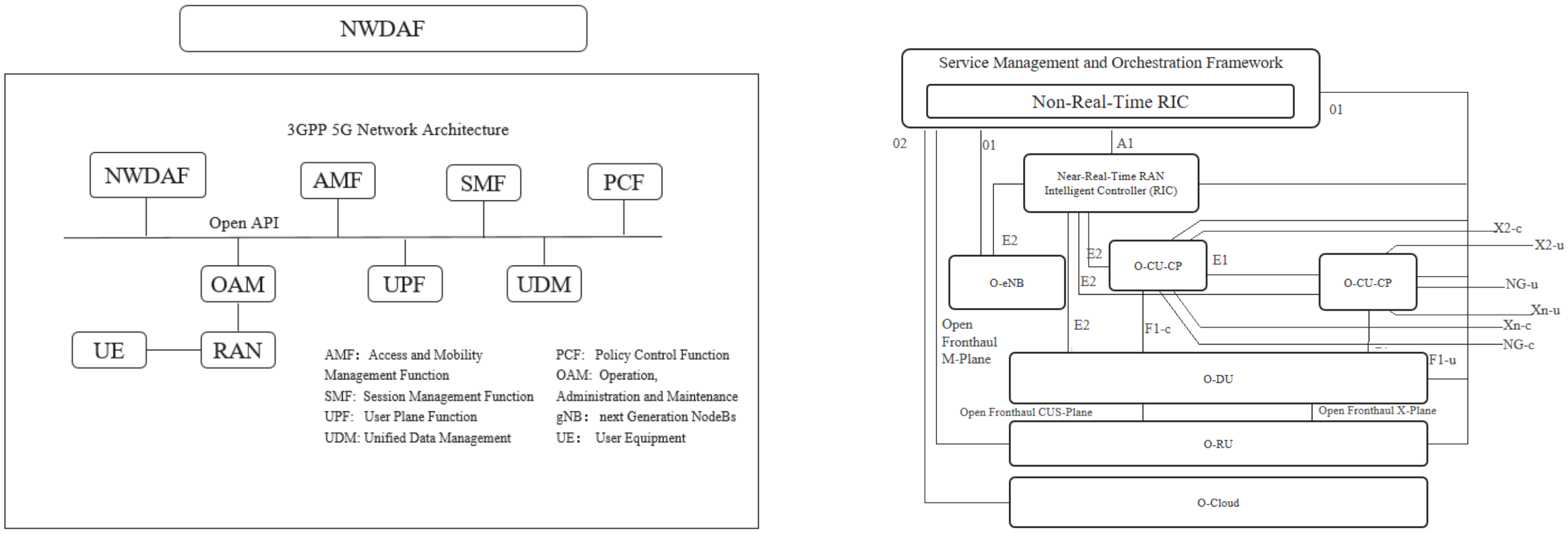

(a) NWDAF framework (b) RIC framework

Figure 2 NWDAF and RIC Framework

With respect to 3GPP, O-RAN is another emerging technology track. Since 2018, O-RAN Union has formulated AI-enabled RAN intelligent controller (RIC), and interfaced with Management & Orchestration (MANO) function of core network. Therefore, the author expounds the RIC in the core network chapter. RIC framework is shown in Figure Figure 2(b). The RIC is divided into Non-RT RIC (non-real time RIC) and Near-RT RIC(near-real time RIC), in which the Near-RT RIC connects CU/DU at the radio access network side through E2 interface, and its function includes the utilization of AI capacity for the radio resource management, mobile management, wireless connection management, switch management, wireless QoS management, etc.; Non-RT RIC is defined in the core network MANO system and connects the Near-RT RIC through AI interface, and its main function is based on the AI-based service and policy management, high-level service process optimization, help Near-RT RIC train AI models offline, etc. RIC of O-RAN is in the process of standard development and its early trials. With respect of 3GPP NWDAF, O-RAN RIC is a long way from mature commercial application.

- Transport Network

Transport network, as the foundation of communication network, is responsible for the physical connection of various network nodes and data transmission. Since optical communication has the features of high-bandwidth, high-stability, and low-insertion loss, etc., current traditional transport technology conducts communication through optical network carrier. Optical transport network development goes through the development and innovation of plesiochronous digital hierarchy (PDH) [86], synchronous digital hierarchy (SDH) [87], multi-service transport platform (MSTP) [88], wavelength division mulplexing (WDM) [89], automatically switched optical network (ASON) [90] and optical transport network (OTN) [91] technology. In order to achieve the flexible control of network flow and improve the network transmission service quality, the industry has explored to introduce SDN to optical networks in recent years to achieve the software defined optical networking (SDON) [92]. SDON inherits the features of dynamic service interruption recovery of ASON, but also devotes to guaranteeing the network capacity and service reliability. Besides, due to the requirements in reduction of network operation cost and the improvement of automatic and intelligent level of optical networking, the optical networks need to combine big data, AI, cloud network integration, etc. to achieve its intelligent management. The concept of cognitive optical network (CON) and the exploration of industry applications are also introduced [93]. In accordance with the project objective of cognitive heterogeneous reconfigurable optical network (CHRON) funded by the European Union [94][95], cognitive optical network is based on the cognition and decision system (CDS) which is responsible for managing network transport requirements and events, and the control and management system (CMS) which is responsible for network controlling and signaling, as shown in Figure 3. Currently, there are some research results about the combination of SDON/CON and AI, such as fault prediction, the reduction of recovery time, and the improvement of light signal-to-noise ratio [96], etc.

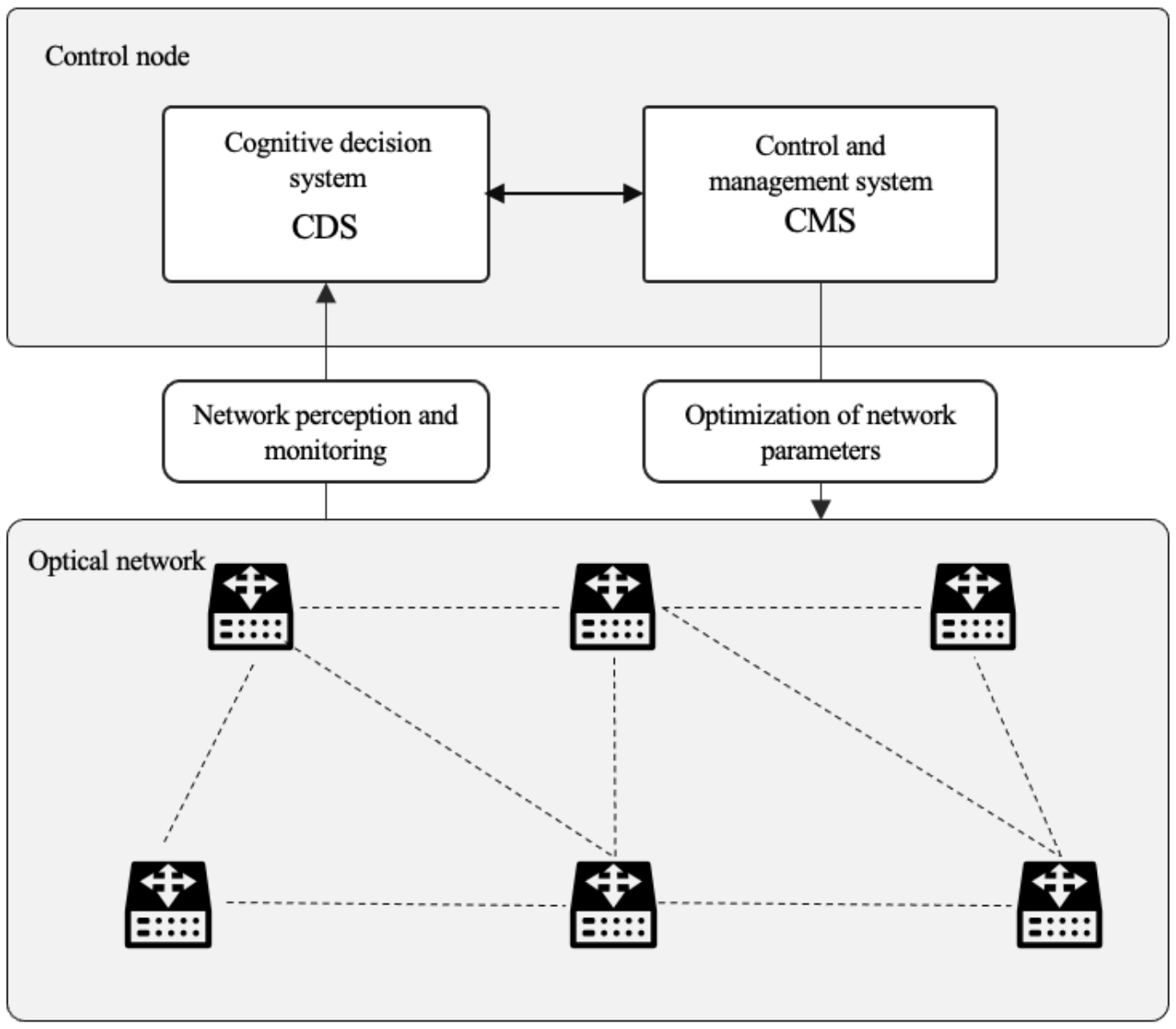

Figure 3 System Structure of Cognitive Optical Network

The evolution of internet protocol version 4 (IPv4) toward the internet protocol version 6 (IPv6) focuses on solving the address space, QoS guarantee of network transmission, etc. In order to meet the requirement of 5G scenarios, the construction of IPv6-based intelligent IP network becomes the development trend of bearer network. Flexible network routing, network slice service level agreement (SLA) of bearer network and deterministic network transmission shall be guaranteed with AI technology, for example, the IP network intention can be recognized and judged based on AI technology for targeted guarantee of network experience. But as a whole, the current combination of IPv6 network and AI is at the initial exploratory stage. The industry hopes to monitor the operating state of the whole network through AI technology in order to promptly find the network problems and risks and intelligently recognize the network abnormality; The cause positioning is conducted for fault and the relevant optimal strategy is generated to solve the problems found. In order to achieve the intelligent IP network better, the segment routing IPv6 (SRv6), in-situ flow information telemetry, etc. need to be introduced [97][98] for the network perception and flexible route configuration, and the IPv6 technology is upgraded to IPv6+. Currently, the relevant research in industry is still at the exploratory stage.

Currently, the resource allocation, storage and computing power of cloud, network, edge and terminal are relatively independent, for example, the complex AI application is conducted at cloud, and the simple and lightweight AI application is conducted at terminal. With the development of SDN, IPv6, IPv6+ and other technologies, the industry devotes to achieving the integration of computing network and IP network, cloud network integration, and other new architectures. Currently, there are many technical problems in the process, such as how to achieve the optimal route, how to distribute computing power optimally and how to guarantee the service quality of computing power. All of them shall be overcome with the help of AI, etc. Now, the relevant research is still at the initial stage.

- Terminal

Terminal-based AI includes the intelligentization of terminal and chip. The terminal operation system and the App on application layer have obtained some intelligent application development. The empowerment of network infrastructure from terminal-based AI is at the early development stage at present. Relatively typical application is to report the performance data of terminal chip acquisition to the SON system or operation supporting system (OSS), and the network AI analysis engine in those two parts is utilized for intelligent optimization of wireless network. According to the 3GPP standard, the intelligent optimization is reflected through the minimization of drive tests (MDT) in 3GPP SON, which has been introduced in Section 4.2 radio access network.

As a whole, the AI in communication network infrastructure field has relatively plain development in 3G/4G era, while it achieves the accelerated development in 5G era. With the gradual maturity and accelerated commercial progress of NWDAF of 3GPP, RIC of ORAN and RAN data analytics function (RAN-DAF) of 5G Mobile Network Architecture (5G-MoNArch) project group[99], AI will further integrate into network architecture of 5G and B5G deeply, exists and evolves for a long time in the form of independent network element and network function entity. In the meanwhile, the AI mathematical model and mobile communication knowledge are relatively independent. Many mathematical model results lack the explanation with physical significance on the communication layer. They shall be further integrated deeply to enhance the interpretability of AI in communication physical network application. The traditional communication equipment manufacturer also needs to open data interface of black box in terms of standardization and helps operators to construct neutral intelligent network infrastructure.

### 3.2 Development of Telecommunications AI in Network Management

The development of AI in network management is described in three parts: MDAF, ENI engine and network operation support system OSS.

- MDAF

The MDAF defined in 3GPP R16 and the service object are shown in Figure 4(a). 3GPP SA5 starts defining intelligent function of network management in R16, for example, the management data analysis function MDAF conducts the data analysis to help management system set rational network topology parameters for network configuration and guarantee service quality. After the rational configuration of network according to the analysis results provided by MDAF, the control plane and user plane can conduct further parameter adjustment to improve user experience. It is critical for the management system of OAM (operation administration and maintenance) to provide business requirement analysis information for MDAF, for example, communication service management function (CSMF) of network slice translates the customers' SLA into communication service demands, judges whether such demand matches with the existing slice case by utilizing the analysis capacity of MDAF, and selects the optimal slice case for slice SLA guarantee [100].

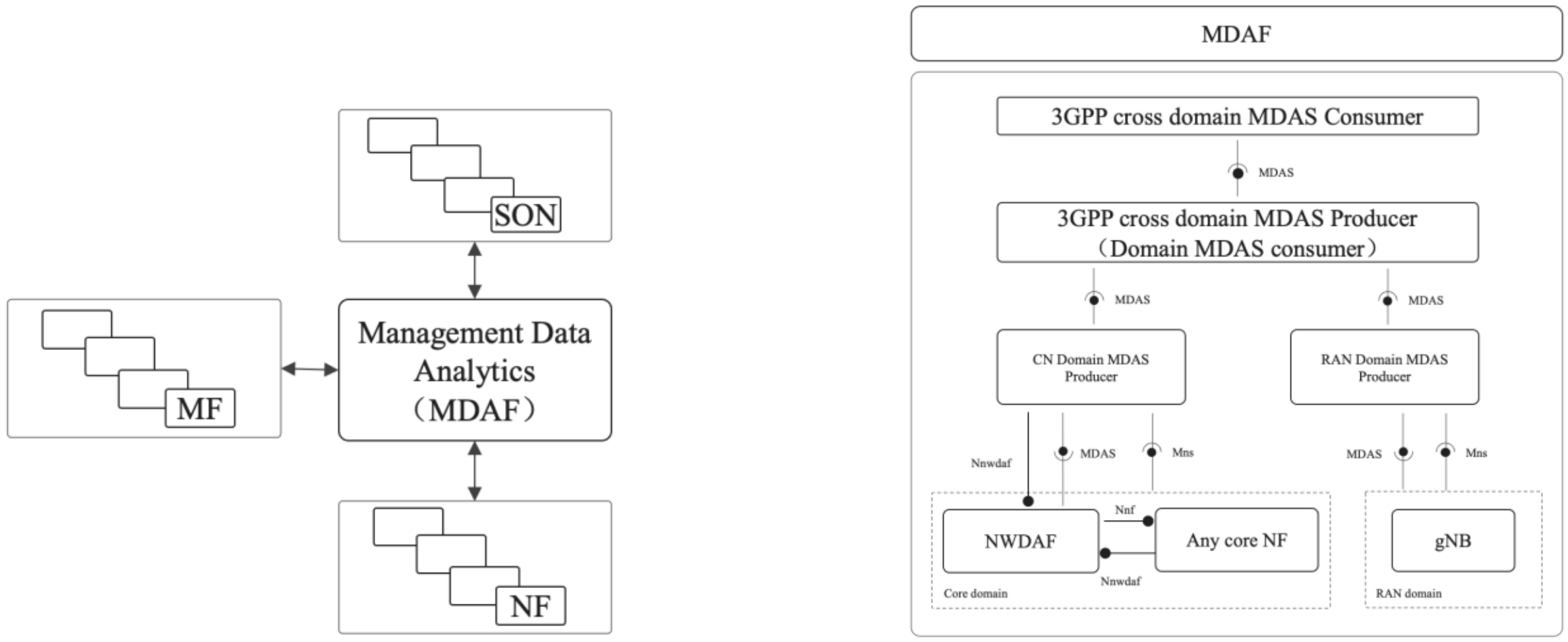

(a) MDAF defined in 3GPP R16 and the service object

(b) MDAF architecture defined by 3GPP R17

Figure 4 MDAF defined in 3GPP R16 and the service object [101] and MDAF architecture defined by 3GPP R17 [101]

MDAF can also empower SON on management plane, as shown in Figure 4(a). MDAF utilizes the collected management plane and network data for the relevant analysis to achieve various SON functions described in Section 3.1. But MDAF standard (such as the definition of interface, collected data information, process, etc.) in R16 is imperfect and is relatively difficult to apply and deploy in current 5G network. In order to solve problems of R16 and face new scenario, 3GPP starts enhancing MDAF in R17 by defining and improving the coverage enhancement, resource optimization, fault detection, mobile management, energy-saving, paging performance management, SON cooperation, etc., in addition to improving the function in R16 [101]. The relationship between MDAF as defined in R17 and service object is shown in Figure 4(b). MDAF has a more complex service framework in R17, and hasn't determined how to coordinate with NWDAF. Its technical standard in R17 is in progress. MDAF hasn't achieved commercial deployment in the 5G network management system of the Chinese and American operators.

- ETSI ENI

ETSI defined the ENI system in 2017. As an independent AI engine, it provides intelligent service for network operation and maintenance, network security, equipment management, service orchestration and management, etc.[35]. ENI functional framework is shown in Figure 5ENI contains the AI-related knowledge management, model management and strategy management module such as context perception, knowledge management, cognitive processing, context awareness, model driving, strategy management, etc. The raw data are conducted cleaning and feature processing through input processing and normalization, then the relevant strategy or directives are provided for the OSS, BSS, user, system application, orchestrator, infrastructure and other service objects after the internal AI module processing. Denormalization, output generation and other modules translates the output strategy or commands of ENI and outputs the language that the service object can understand.

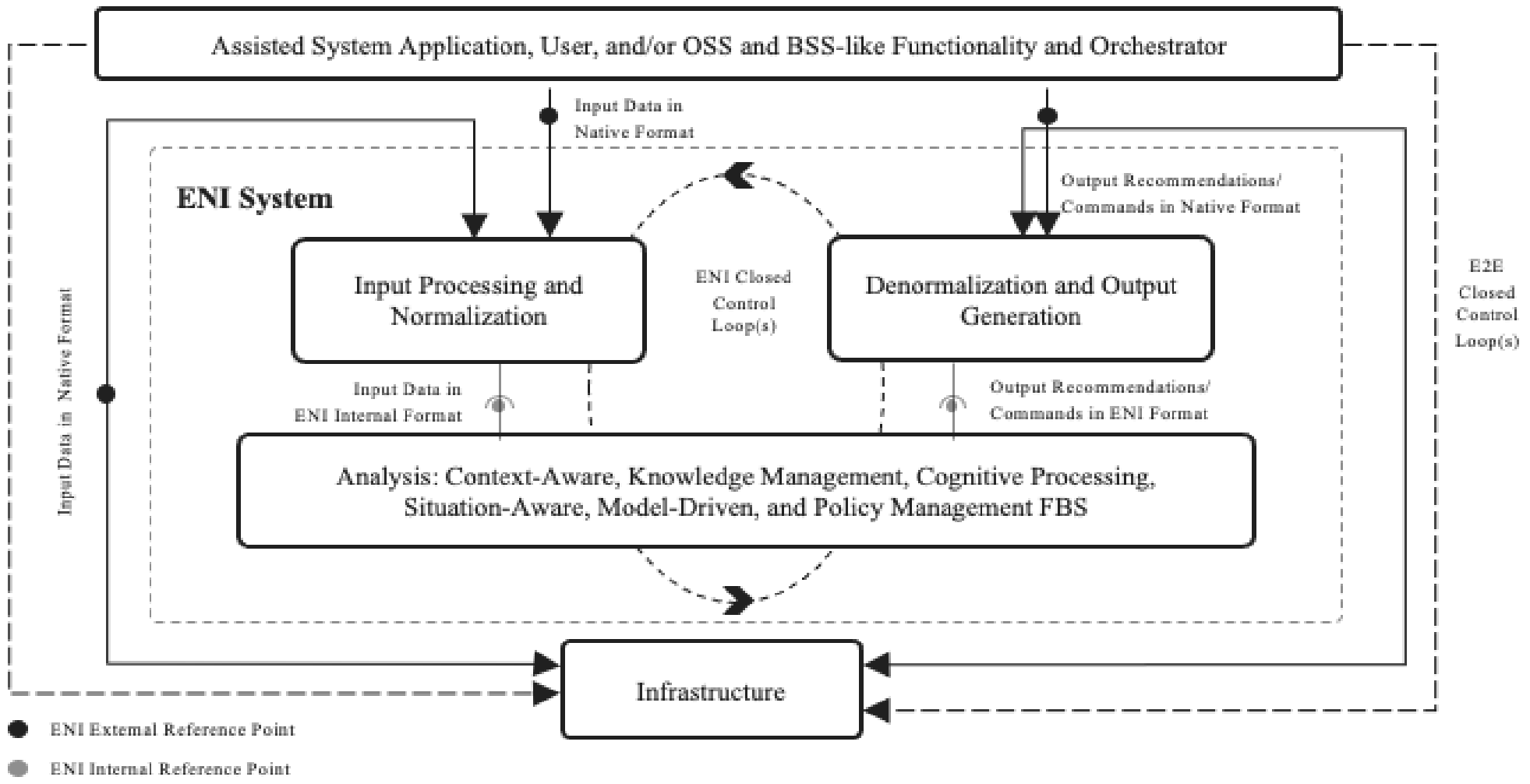

Figure 5 ENI Functional Architecture [102]

Currently, ETSI ENI continuously evolves the function of ENI, and defines more senior applications, such as the energy saving based on intent based network [102], data mechanism [103], matching of ENI and operator system [104] etc. The relevant work is still ongoing. For the function and application scenarios defined by ENI, the domestic and foreign operators have attempted the relevant prove of concept (PoC) project, and have obtained favorable effect in slice management, user experience optimization, wireless energy optimization, etc. [105]. The PoC report displayed in the references [106] shows that UC3M, Samsung, Telecom Italia, etc. cooperate the flexible control of slice resources through ENI to control the end-to-end delay, service creation time, system capacity, etc. The report displayed in references [107] shows that China Telecom, AsiaInfo, Beijing University of Posts and Telecommunications, etc. cooperate to optimize the intent-based user experience of ENI to improve user experience management. Currently, ENI hasn't been arranged in 5G network or network management system as an independent AI system or network element, but many functions defined by it are applied in network management system of global operators through the decoupling method.

- Network Operation Support System

The network Operation Support System (OSS) is often defined as a kind of software function so that the communication operators can manage its network and application. An OSS system often has five functions at least: network management system, service delivery, service fulfillment (including network inventory and network activation and provisioning), service assurance and customer care [108]. AI application was also developed very slowly in the OSS domain at early stage. In the 1970s, most work of OSS was conducted manually and intervened manually; In the 1980s, with the emergence of Unix system and C language, Bell System started developing OSS, and the famous OSS at early stage includes automatic message accounting tele-processing system (AMATPS), centralized service order bureau system (CSOBS), engineering and administrative data acquisition system (EADAS), switching control center system (SCCS), service evaluation system (SES), etc. AI wasn't applied in those early systems. Till the 1990s, ITU-Telecommunication Standardization Sector (ITU-T) defined four layers of new frameworks of OSS in telecommunications management network (TMN): business management level (BML), service management level (SML), network management level (NML) and network element management level (EML). Later, FCAPS (fault, configuration, accounting, performance, security) as a new model, was introduced to manage such 4 layers. In the commercial process of TMN, AI was also rare. After 2000, with the development of NGOSS (new generation operations systems and software) project and enhanced telecom operation map (eTOM) framework of TM Forum, OSS framework system of communication operator has relatively standardized guidance standards. The application of AI also has a relatively stable framework and carrier, especially for the fault management and performance management in FCAPS, AI starts attempt and application in the fault diagnosis, root cause analysis, performance prediction, etc. [109].

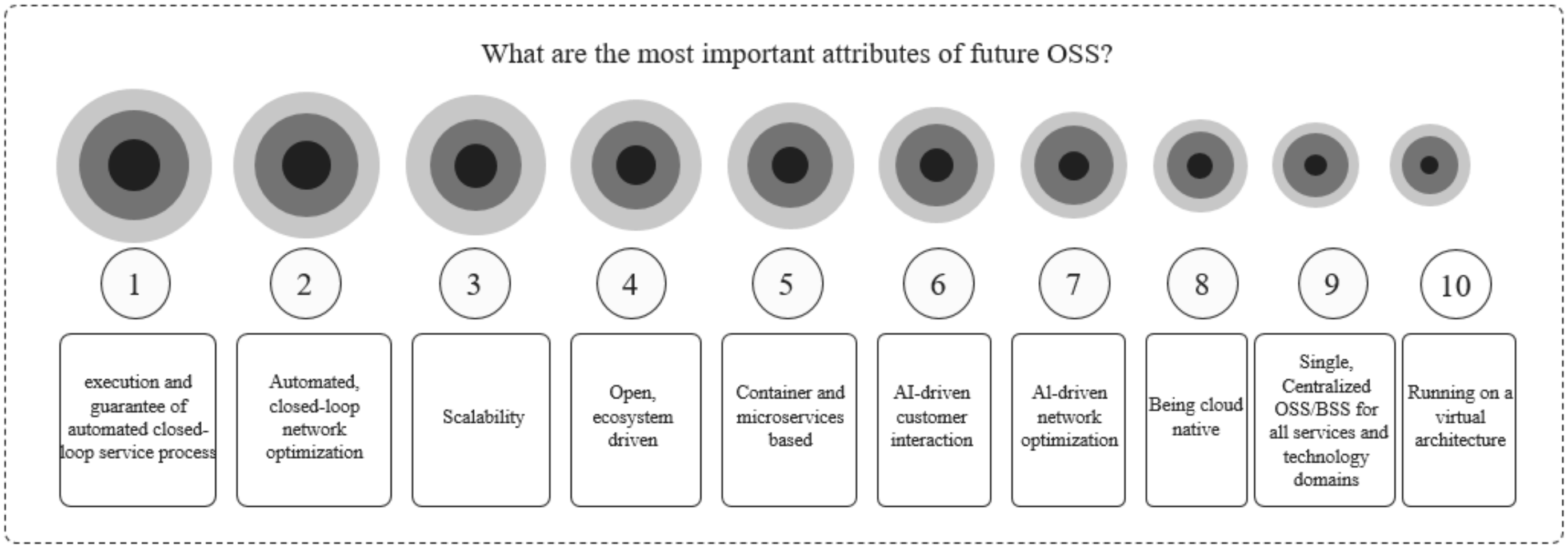

Figure 6 OSS Attribute Defined by TMF

OSS attribute defined by TMF is shown in Figure 6In 2019, TMF defined the data-driven future OSS, which must depend on the tight coupling of AI, machine learning, automation, micro-service and service optimization and must be agile, automatic, proactive, predictive and programmable in the research report about Future OSS [110]. Among 10 most important factors to define future OSS, 4 factors are closely related to AI: execution and guarantee of automated closed-loop service process, optimization of automated closed-loop network, AI-driven customer interaction and AI-driven network optimization. Therefore, mainstream communication operators gradually embed AI platform or function module into OSS system for 5G evolution to expect the intelligent evolution of OSS. In the enhanced control, orchestration, management & policy (ECOMP) system of AT&T [111], the AI analysis-based Analytic Application Design Studio function is defined at the design state, as shown in Figure 7 In the operation execution state, ECOMP also defines the function of data collection, analytics and events (DCAE) which provides the AI-based real-time FCAPS function and achieves automated closed-loop goal through the intelligent analysis and auto-orchestrations of services, network and resources respectively [112]. Verizon also deems AI as required function of BSS/OSS, and AI is deeply applied in the user network and service experience [113].

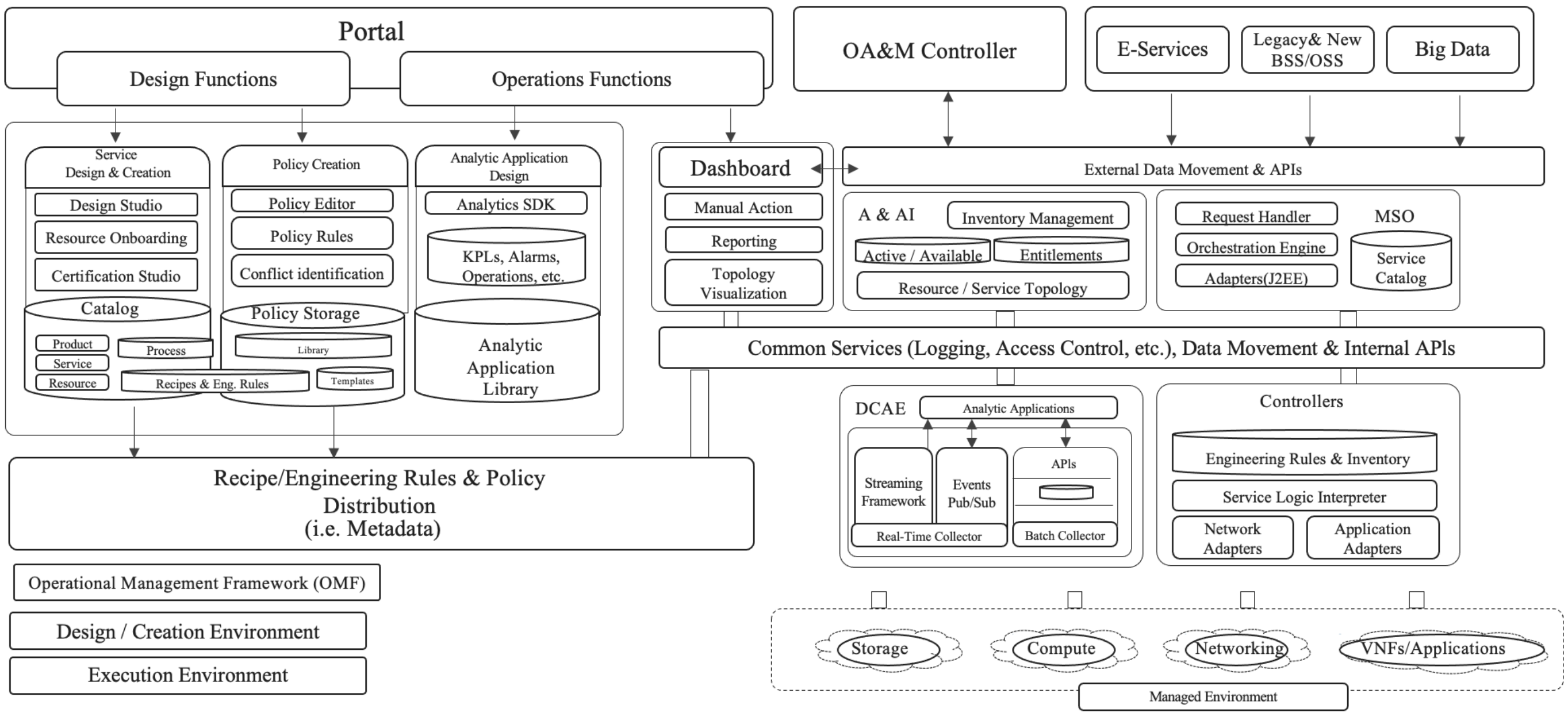

Figure 7 ECOMP defined by AT&T

China's operators are observed to insert a new platform or component between data middle office and OSS core function module in the face of the system construction of 5G OSS, which is named as network AI middle office or intelligent middle office to undertake the network AI function. In Figure 8(a), the author abstracts 5G OSS network middle office system of China's three communication operators to work out an architecture diagram of 5G OSS network middle office with neutral technology, in which the data middle office is mainly responsible for the data acquisition on the network side, data storage, data governance, data sharing, etc. In order to meet the intelligent requirements of 5G network, service management, the network AI middle office adopts the data middle office big data on the internet as main resources and centers on the planning, construction, optimization, maintenance and other scenarios in network lifetime to continuously construct, reason, release and accumulate network AI algorithm model, and provides the 4G/5G network with the network AI function including anomaly detection, capacity prediction, network optimization, root cause analysis, alarm prediction, fault self-healing, service orchestration, perception optimization to comprehensively boost the automation and intelligentization capacity of 5G network.

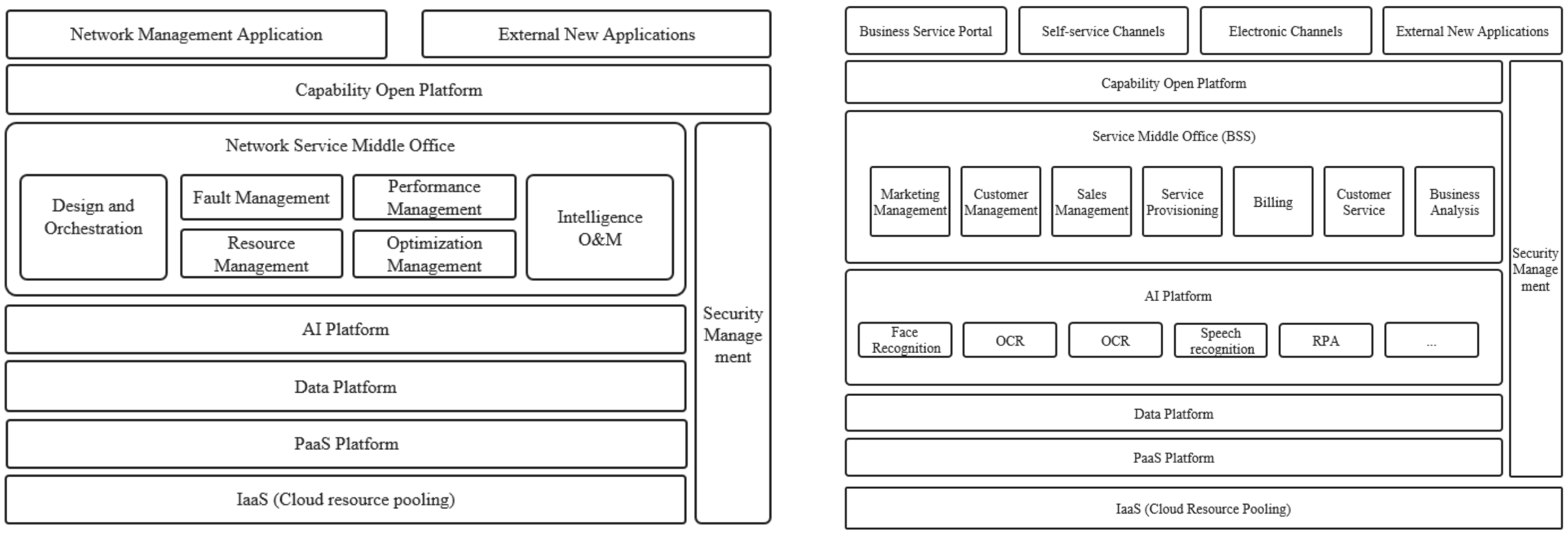

(a) 5G network operation support system (b) 5G business support system

Figure 8 5G network operation support system and 5G business support system

## 3.3 Development of AI in Telecommunication Service

Communication BSS provides customer operation and service. In the system framework of TMF, BSS mainly focuses on customer marketing, telecom products, customer service, resource management, supplier management, partner management, etc. [113], as shown in Figure 8(b). The core of BSS includes customer relationship management (CRM), billing system, business intelligence (BI) system, call center (CC) system, etc. The traditional BSSs in all over the world have been mature, and most of them have completed the construction of centralized platform. Chinese telecom operators are leading the technology evolution based on intelligent middle office in BSS system, namely accomplish the IT service for users and partners by telecom cloud service operation system, which includes service middle office, data middle office, technology middle office, AI middle office, etc. The AI middle office, based on AI algorithms, empowers the telecom service with intelligence through the scenario-based service operation. Currently, AI technology has been applied well in various BSS services, including marketing, customer experience management, customer service, and billing through AI middle office system.

- Marketing

The typical AI application in telecom marketing is product recommendation and marketing operation decision assistance. The product recommendation is usually designed on various AI models and marketing expert knowledge, and the optimal marketing strategy is implemented based on the AI strategy model which includes various customer features, for example, product-customer matching, customer high-value potential, for a comprehensive marketing decision and enterprises income improvement. The AI applications include: recommendation of popular products, recommendation of relevant products, recommendation of personalized package, recommendation of agreement, recommendation of digital content, etc.

- Business Sales

The typical AI application in this field is the use of face recognition, Optical Character Recognition (OCR) and other technologies to support the Service Hall customer services, such as identity authentication audit, certification of sales agreement signature, and confirmation of real person business transaction. In the sales process of government and enterprise customers, OCR and image recognition technologies are used to support the automatic identification of enterprise information, seal identification of government enterprise business for prior authentication auditing, the automatic drafting of sales contracts and other scenarios. These AI technologies achieve the intelligentization and automation of business sales and boost the working efficiency of customer managers.

- Customer Service

The typical AI application in this field is to achieve the voice interaction between customers and intelligent robot and the real-time monitoring of customer emotion, prediction of customer demands, effective distribution of service seat, real-time monitoring customers' problems, automatic classification and recognition, automatic retrieval of knowledge library, assistance of seats in replying questions, identity authentication based on customer voiceprint, prediction of potential complaints, voice quality inspection in customer service process, intelligent quantitative scores, intelligent bill dispatching based on the work sheet text information, automatic generation of knowledge in knowledge base, intelligent customer service scheduling, etc. based on the voice recognition, intention recognition, multimodal Q&A matching, voice synthesis, semantic processing, user sentiment analysis, label multi-classification prediction, OCR, etc.;

- Billing

The typical AI application in this field is to support the upgrade and configuration of billing system application process and the amendment of online gray release by using artificial intelligence for IT operations (AIOps) to achieve the fault discovery, fault diagnosis, fault self-healing, fault prevention, etc. of billing system; The automation of daily account book registration of billing system is supported in combination with the robotic process automation (RPA). Besides, based on the multidimensional quantity pricing factors (bandwidth, time delay, reliability, precision, linking number, capacity, number of network function instances, etc.) and customer data, the intelligent pricing can be achieved through AI algorithm to conform the optimal price and help B2C enterprises obtain the maximum benefits.

### 3.4 Cross-domain Intelligent Integration Development of Telecommunications AI

Any telecommunications service can't separately operate or maintain in the single system of OSS or BSS. The integrated telecommunication business, the evolution of technology middle office, and integrated data analysis of telecom service and communication network are deemed as three main factors to drive the deep integration of BSS and OSS system. Telecommunications AI also generates many application scenarios and use cases in the development of cross-domain intelligent integration.

- Customer Experience Management

In the organizational structure of telecom operation services, the network operation support and the telecom business support are relatively independent domains, and their corresponding production system OSS and BSS are also relatively independent in operation and evolution. The telecom network domain focuses on various key performance indicators (KPIs) of network communication and network functions, while the telecom business domain is responsible for the market-oriented development of new customers, new business and inventory customer service. Telecom operators often use the QoS system, formulated by ITU and ISO, as the SLA with customers. Traditional QoS is driven by technology and defines QoS with network KPIs. It can't truly reflect the customer experience and emotion of network usage. Therefore, there is a "digital gap" between the network performance quality based on QoS system in the network domain and the customer satisfaction perception in business operation domain [114], as shown in Figure 9. Customer Experience Management (CEM), which is a new field of cross-domain integration of network and business, evolves the QoS system of operators toward the user-centric QoE system by AI technology to achieve the network business service transformation from centering on network KPI to centering on customer experience [115].

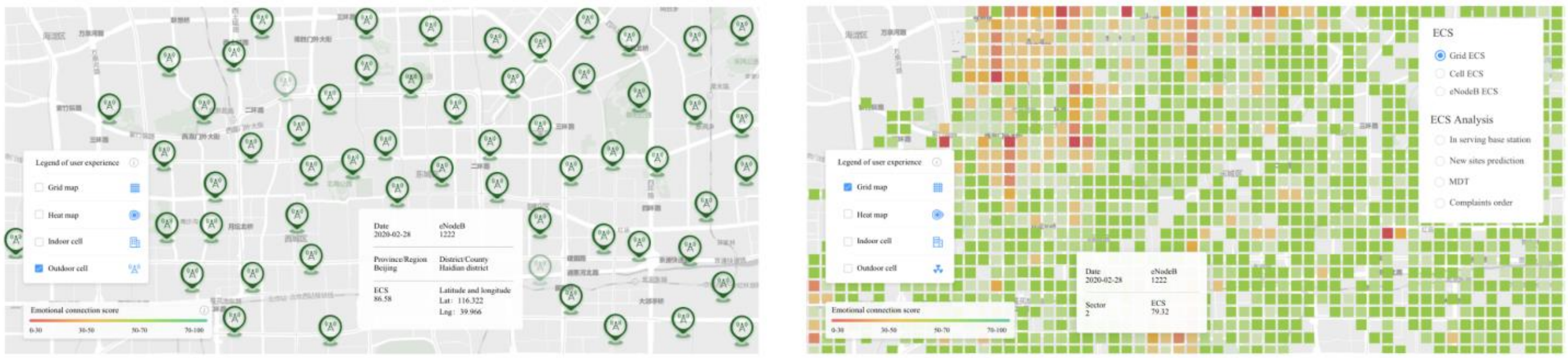

Figure 9 Gap between Performance and Experience

The core of CEM is to establish a set of AI models which can accurately perceive the customer experience about the use of communication network and telecom business through AI algorithms in combination with psychology, which is called as telecom psychology algorithm. Such algorithm conducts the quantitative mapping of network system's QoS and QoE to remedy the gap between network quality and users' real experience. Currently, there are two kinds of methods to evaluate user experience perception: net promoter score (NPS) and emotional connection score (ECS) [115][116], and as shown in Figure 10, NPS is used to measure the possibilities whether customers recommend certain company, product and business [116][117]. NPS constructs the scoring indicator system based on customer feedback, interviews survey users by telephone or questionnaire to quantize the satisfaction of certain company, certain product or service between 0~10 scores [116]. NPS is a kind of passive and static quantitative evaluation mechanism based on customers' long-term impression. Now, about 7% global telecom operators use the NPS to measure the customer satisfaction [118].

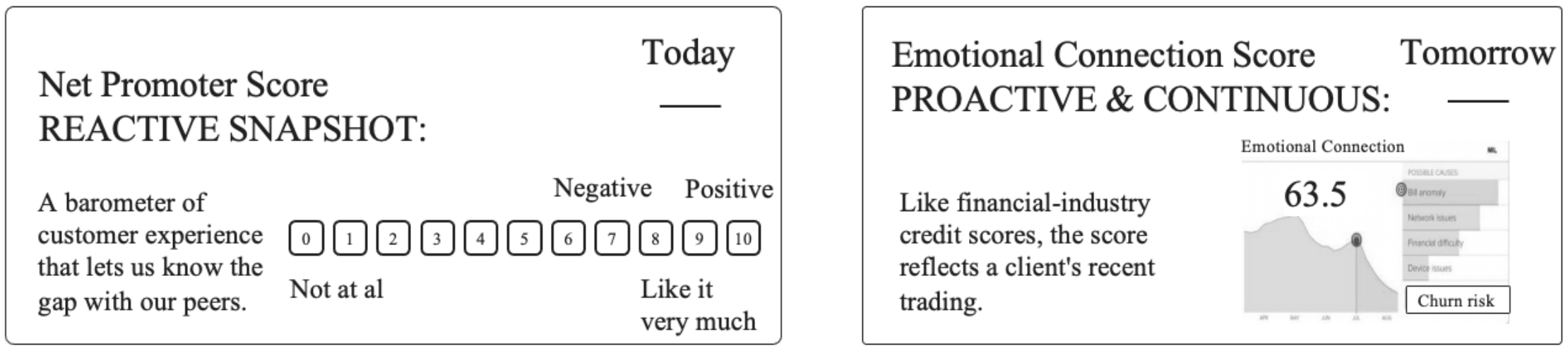

Figure 10 Indicator System Comparison: Net Promoter Score (NPS) and Emotional Connection Score (ECS)

As the ECS model about machine learning integrating psychology occurs, traditional NPS indicator system has been out of date. Lots research results find the emotional connection score of customer is the index system which is closest to the true experience quality [115]. Different from NPS, the ECS is proactive and persistent. As shown in Figure 11the relationship between quantizable customer experience and various indicators can be continuously learned through the ECS telecom psychology algorithm by converging and accessing various data across network and business domain, and event results are mapped to ECS scores to promptly explore the root causes of customer experience problems, guide to improve network and service quality and boost user experience. AsiaInfo proposes a set of telecom psychology experience perception algorithm and indicator set of quantifiable user perception experience [116]. The massive data machine learning is conducted for the user-level subjective data (such as NPS survey, customer complaint, active dial testing, etc.) and objective data (such as voice communication, surfing the internet, HD video business, VR business and other quality indicators) in communication field to compare and verify user data difference among different regions for the ECS parameter optimization, then the user characteristics are analyzed in combination with user-level communication, consumption, service, etc. to finally generate the ECS telecom psychology model which is used for real-time evaluation of instantaneous experience quality of any service at any moment of customer journey and in any place.

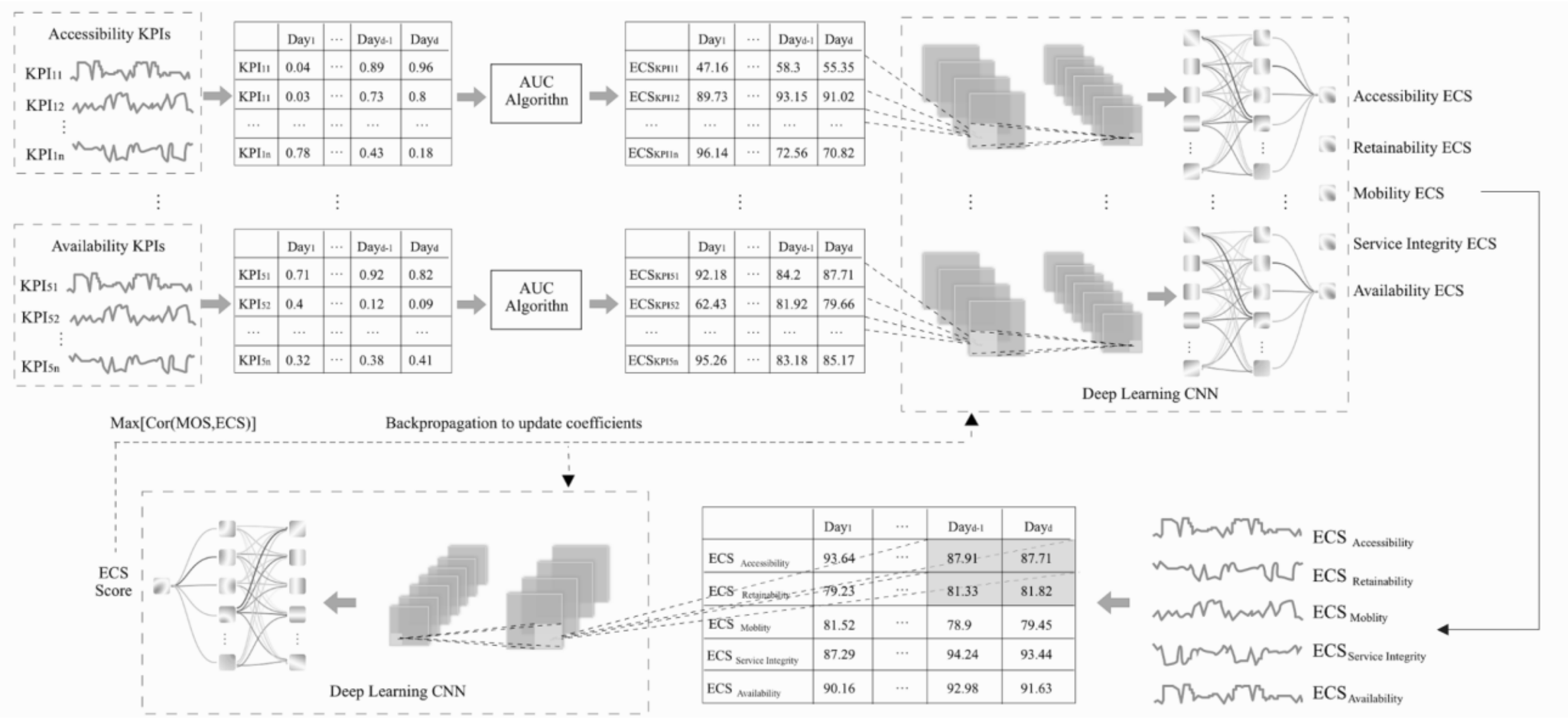

Figure 11 ECS Telecom Psychology Model

With the full life cycle journey of customer shown in Figure 12 the application of telecommunications AI in CEM aims to improve user experience, evaluate the instantaneous experience quality of service of users, conduct rapid positioning and diagnosis of experience reduction, and penetrate the user personalization strategy about proactive perception and care into it, thus clearly knowing the experience indicator of every user in network journey. The personalized service is provided for users through network and business system to achieve the 5G network personalization (NP).

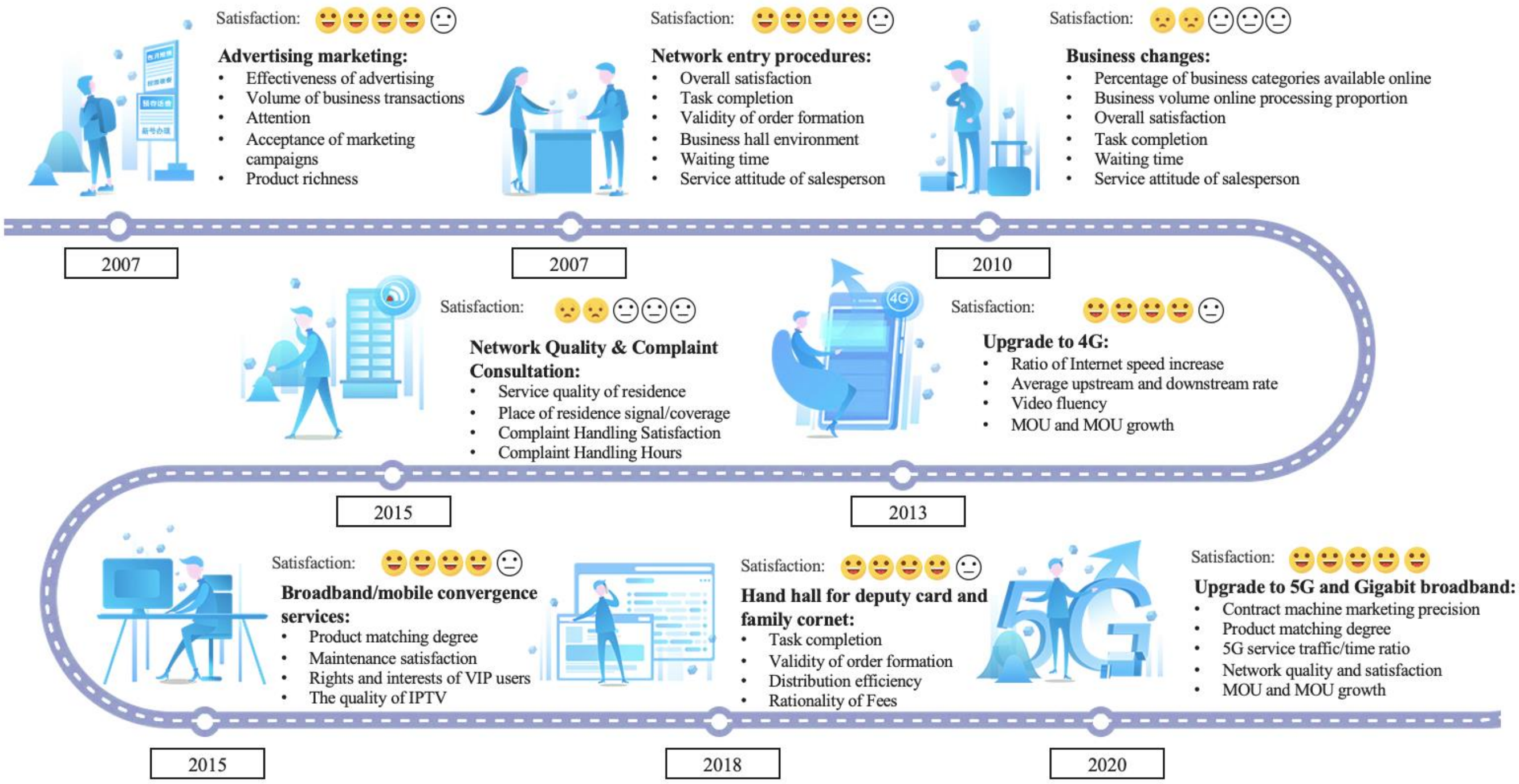

Figure 12 Full Life Cycle View about Customer Experience

- PCF+ (policy control function+)

Network policy control is mainly based on rule definition, and is achieved through policy and charging rules function (PCRF) in 4G LTE [119]. 3GPP introduces PCRF network element from R7 (Release 7) to control the user and service QoS, and provides differentiated service for users. Besides, it can provide users with service flow bearing resources guarantee and flow billing strategy to achieve finer service control and billing method based on service and user classification and rationally utilize network resources. PCRF contains strategy control decision and the function based on flow billing control, provides network control function of data flow detection, gate control and flow billing, and can generate control rules by triggering many dimensions such as business, user, location, accumulative usage amount, access type, time, etc. AI isn't applied in PCRF, and the PCRF's strategy rule is mainly based on rule configuration. PCRF framework, as shown in Figure 13(a), executes strategy and billing function through interaction with network elements such as GX reference point, policy and charging enforcement function (PCEF), etc.

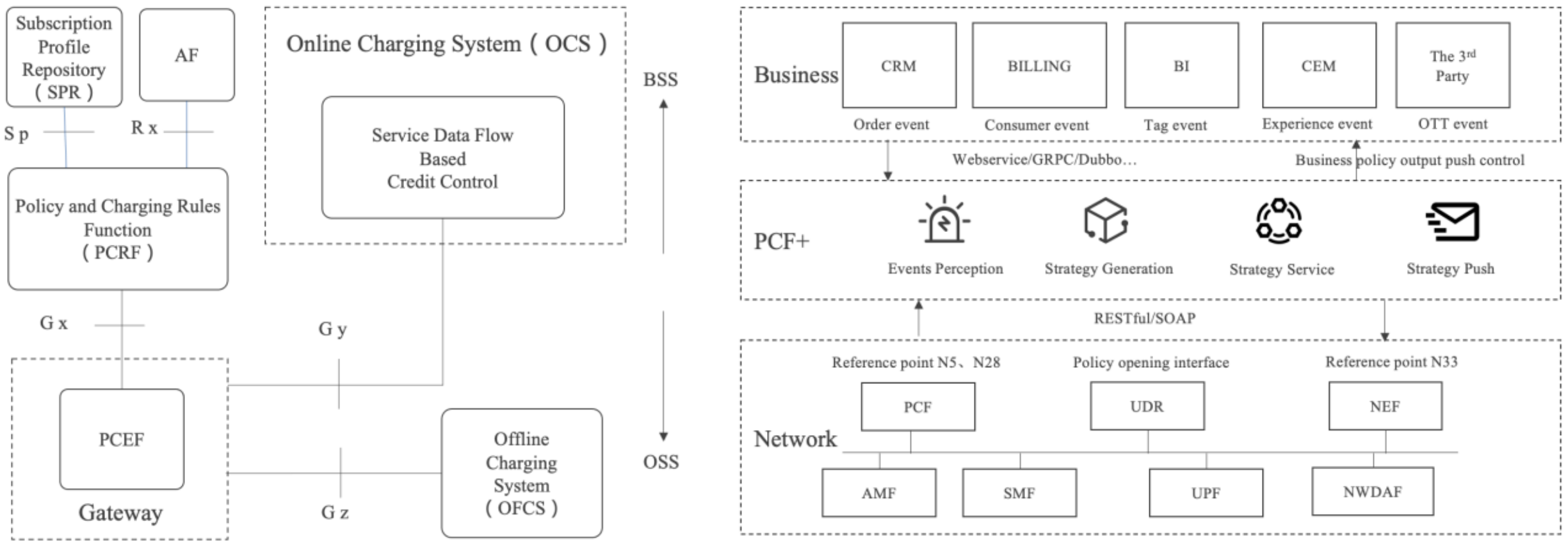

(a) 3GPP PCRF framework (b) PCF+ based on BO domain integration

Figure 13 3GPP PCRF Framework and PCF+ Based on BO Domain Integration

3GPP only defines the PCF for network side, but still needs to enhance function by combining BSS and OSS (BO) domain because users propose higher requirements to 5G service differentiation with the enriching of 5G service. In addition to proposing higher requirements to network resource use and network control, operators still need to own refiner analytical capacity combining business strategy. Such analytical capacity often requires to master the user attribute characteristics or event information in business domain, which requires expanding the action range of PCF, collecting data in both OSS domain and BSS domain, and expanding application objects from network domain to business domain. Based on the above consideration, PCF shall evolve to PCF+ to provide new business mode, business scenario or business model, Figure 13(b).

Besides, the introduction of AI / big data capacity to 5G core (5GC) causes the strategy control becomes more intelligent, instead of being based on expert system or rule configuration function like previous PCRF. The integration of NWDAF and PCF/PCF+ can easily achieve the intelligent slice experience management, intelligent SLA guarantee, etc. Besides, PCF+ can utilize data (obtaining the cell congestion status from network domain and obtaining the user level and package use condition from BSS domain) of network domain/business domain integration and combine the AI technology to conduct dynamic adjustment of user package setting or recommend the optimal package for the user's QoS guarantee and the improvement of user satisfaction.

## 3.5 Development Case of Telecommunications AI in Private Network

5G not only provides public network service for vertical industry, but also can provide private network service. Telecommunications AI can provide a series of intelligent private service and security guarantee in 5G private network. According to the sharing model of network resources, the 5G private network deployment is divided into 3 types: virtual private network, mixed private network and independent private network, as shown in Figure 14in which the virtual private network, based on the current 5G public network, achieves the service bearing of professional users in slice method and has the following features:

- Virtual private network and public network share the UPF network function;
- Private network can connect to enterprise through existing virtual private network in combination with customers' requirement; Mixed private network, based on the control plane of 5G public network and broadband wireless access system, bears private network service, with the following features:
- Support the privatization of user plane data, independent UPF and multi-access edge computing (MEC), with equipment established in park;
- Conduct the access authentication of users through slice identification.

Independent private network and operator 5G public network are completely separated, with the following features:

- Adopt the simplified core network and achieve the independent networking in combination with UPF and wireless base station;
- Special base station is covered without dead angle as guaranteed, UPF/MEC guarantees the accumulation of private network data and the network independence.

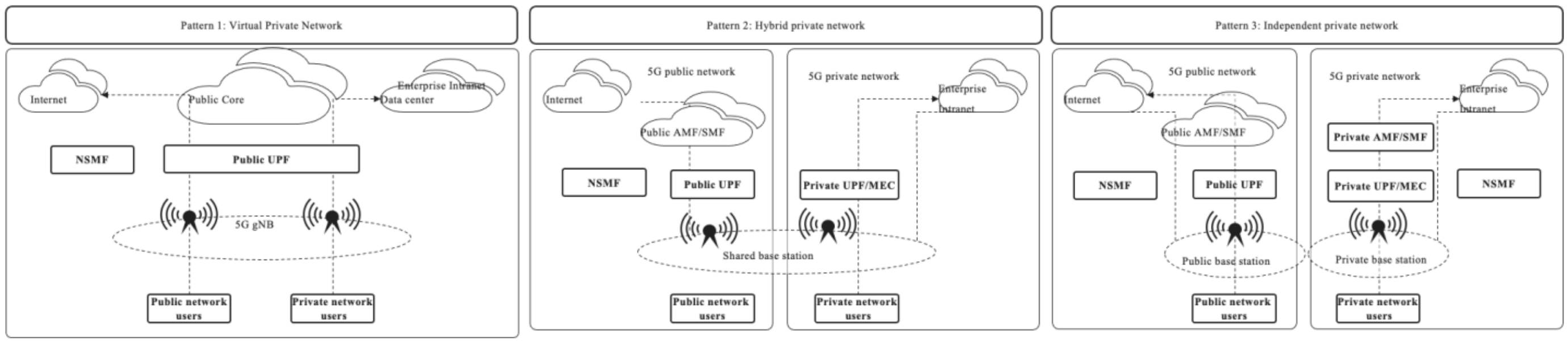

Figure 14 3 Kinds of Models of 5G Private Network

In the above 3 kinds of modes, telecommunications AI can be applied in any private network mode, for example, it can be used for network slice SLA guarantee in virtual private network to conduct optimization of transmission performance, quality and resources. For independent private network, the AI experience perception algorithm can conduct real-time or quasi real-time evaluation of the users' perception and experience, and provide accurate QoS portfolio guarantee service quality to achieve the differentiated intelligent operational service. In private network, the Federated Learning, Transfer Learning and other AI technologies can be utilized to accomplish the cloudification management and continuous learning optimization of 5G slice abnormity diagnosis model. Besides, AI technology can also conduct real-time evaluation of the wireless network performance in private network [37], adaptive adjustment is conducted for the parameter setting on application layer through interaction with the application platform in vertical industry to boost the video quality on application layer or conduct the game acceleration.

## 3.6 Development of Telecommunications AI in International Standards Organizations

In view of the rapid development of AI technology in 5G commercialization history, international standards organizations of communications also start preliminary grading of development maturity of telecommunications AI. Intelligent automation can't be accomplished in an action, and needs to be conducted step by step. Complete intelligent autonomous network is the ultimate goal, and needs to start from automatic repeated operation means; the automation of network operation is achieved preliminarily at first, then the network environment and status are perceived proactively, and machine learning is needed for making the decisions for continuous optimization; and then user intention is recognized by developing from network perception to recognition to construct the closed-loop cognition learning network, and finally the closed-loop autonomous network from perception, cognition to precognition is achieved, with continuous self-optimization and evolution. For the intelligent grading system of the specific communication network, GSMA, ETSI, TMF, etc. conduct the relevant definition and recommendations [42][120][121], comparison of respective grading standards is shown in Table 1.

Table 1 Internet Intelligent Grading Standards Defined in GSMA/ETSI/TMF

| | GSMA | ETSI | TMForum |
|---|---|---|---|
| Level 0 | - System provides auxiliary monitoring function<br>- Manual execution of dynamic task | -Completely manual | -Manual O&M |
| Level 1 | -Subtask execution according to the existing rules | -Network management system generates batch of equipment configuration script | - Auxiliary O&M |
| Level 2 | - Enabling closed-loop O&M for some units | -Achieving partial autonomy | -Part of autonomous networks |
| Level 3 | -Perceive real-time environment change based on L2 function<br>-Optimize and adjust itself in some fields to adapt to the external environment | -Conditional autonomy<br>-Achieve automated management at some stages of service life cycle | -Conditional autonomy network |
| Level 4 | -Achieve the prediction of service and customer experience-driven network based on L3 function in more complex cross-domain environment<br>-Proactive closed-loop management | -High degree of autonomy<br>-Achieve the business perception, proactive O&M, and autonomous network driven by self-healing and business based on SLA | -Network with high degree of autonomy |
| Level 5 | -Own the closed-loop automation function across service, domain and the whole life cycle<br>-Complete autonomous network | -Complete autonomous management | -Complete autonomous network |

In accordance with the review of telecommunications AI of Section 3.1- 3.5 in each communication ecosystem and in combination with the L0~L5 grading system, the author makes the grading evaluation (as shown in Table 2) for the current application of telecommunications AI in network infrastructure, network management, communication service, cross-domain integration and vertical industry.

Table 2 Application Grade of Telecommunications AI in Telecom Ecosystem

| | Mobile Network Infrastructure | Network Management | Telecom Service | Cross-domain Integration Intelligence | Private Network |
|---|---|---|---|---|---|
| Level 0 | ✓ | | | ✓ | ✓ |
| Level 1 | ✓ | ✓ | ✓ | ✓ | ✓ |
| Level 2 | | ✓ | ✓ | | |
| Level 3 | | | | | |
| Level 4 | | | | | |

As a whole, the development and application of telecommunications AI in telecom ecosystem is still at the preliminary stage. In respect of some applications in the network management domain and business support domain , AI has reached the third level, namely partial autonomy, such as the liberalization and self-healing of SON in network, automatic orchestration and gray release, etc. of business process of AIOps in network O&M domain, etc. Network O&M is competent for the big data acquisition, the computing power can be guaranteed through server cluster, and the intelligent application scenario, such as antenna energy saving, fault detection, etc. is relatively clear and owns inherent basis and platform where telecommunications AI has favorable application, with relatively rapid development. AI service capacities such as intelligent customer service, intelligent marketing, etc. involved in the telecom business domain, can horizontally refer to the similar application experience in other industries very well, with rapid development. The intelligent process in other ecological fields is at the stage L0 and L1 most. The AI development of network infrastructure is to be inspected in NWDAF, MDAF, RIC, etc. and also depends on the opening degree for AI to integrate into the network infrastructure architecture in 3GPP, O-RAN, ETSI and other international communication standards in the future. The cross-domain integration development prospect of AI is also to be decided on the basis of tight coupling effect between the AI and the core system such as CEM, PCF, etc. or network element.

# 4 The Next Decade of Telecommunications AI

In the next decade, mobile communication will comprehensively evolve according to the B5G and 6G standards, which is the critical 10 years for deep integration of telecommunication and AI. In combination with the development process of current telecommunication technology standards such as 3GPP, ITU-R, ETSI, etc., it can be predicted that each field of the telecommunication ecosystem will gradually achieve "complete autonomy" of B5G and 6G in the future by stages through the deep integration with AI.

## 4.1 Development Path of International Communication Standards from B5G to 6G

Since 2018, the European Union, Japan and South Korea, the United States and China have started the pre-research of 6G, for instance, the European Union started the Hexa-X project under the leadership of Nokia [122],Qualcomm, Microsoft, Facebook, etc. in the American industry united NextG Alliance for 6G technology research [123]. Ministry of Industry and Information Technology of the People's Republic of China has expanded the original IMT-2020 promotion group to the IMT-2030 promotion group. ITU also started 6G working team Network 2030 [124]. On Feb. 19, 2020, at the 34$^{th}$ International Telecommunication Union Radio Communication Department 5D Working Team (ITU-R WP5D) Conference convened in Geneva, Switzerland, the formulation of the 6G study schedule and future technology trend research report, and the compilation of future vision technology recommendations, etc. was discussed.

Each communication standard organization is still planning the detailed roadmap of B5G/6G. The whitepaper [125][126] and the prediction of standard roadmap of 3GPP and ITU 6G, the period of 2020-2023, namely at the 3GPP R17-R18 standardized stage, is the research stage of 6G technology trend and vision, the period of 2023-2027, namely at the 3GPP R19-R20 standardized stage, is the research stage of 6G frequency spectrum and performance, the period of 2027-2029, namely at the 3GPP R21 standardized stage, is the stage for various countries to submit 6G evaluation results to ITU.

3GPP predicts to start formulating 6G standard in 2025 at the earliest, will accomplish formulating 6G air-interface standard technical specifications at the soonest in 2026-2027, namely the R20 standardized window, and will submit 3GPP 6G standard to ITU in 2029-2030, namely at the 3GPP R22 standardized stage. It can be predicted that B5G/6G will continue evolving and enhancing air-interface protocol in mobile broadband, fixed wireless access, Industrial Internet of Things, Internet of Vehicles, extended reality (XR), large-scale machine communication, unmanned aerial vehicle, satellite access, etc. and will research and formulate the relevant standards of higher frequency band such as new radio (NR) 52.6-71 GHz and Terahertz. Besides, the service scope of 6G communication standard will expand from land to satellite, seabed and underground to truly achieve trinity communication of sea, land and sky. For the vertical industry such as Industrial Internet of Things, etc., the following standard researches will continue and become mature: narrowband IoT, terminal for wearable and video monitoring, integrated evolution of access and backhaul, direct transfer air-interface of 5G and its evolution function, 5G non-licensed frequency band air-interface, location enhancement, intelligent self-organizing networks, communication sensing integration and its evolution function, network topology enhancements, etc., in which some research work has been carried out in 3GPP SA1/SA2 and other working teams [127].

## 4.2 Forward Looking of Telecommunications AI in Mobile Network Infrastructure

The next decade prospect of AI in communication network infrastructure is expounded in four aspects: Radio Access Network, Core Network, Transport Network and Terminal.

- Radio Access Network

In the radio access network, the application scenario of SON is defined clearly by 3GPP, and the machine learning algorithm utilized previously has obtained certain effect in the industrial field. The authors think the development of SON in the wireless access network of B5G would continue accelerating. AI algorithms such as neural network, reinforcement learning, etc. will also gradually replace traditional machine learning's genetic algorithm, evolutionary algorithm, multi-objective optimization algorithm, etc. In SON, the self-optimization and self-healing would have great development potential. 3GPP SA5 and RAN3 also set up 2 research topics: "Study on SON for 5G" and "RAN-Centric Data Collection and Utilization for Long Term Evolution and NR" [128][129]. In addition to inheriting most use cases of the last generation of SON, 3GPP is suggested to define the interface and signaling protocol of NR, 5GC, OAM and 4G system for the next critical action of SON so that OAM can integrate into the 5G network infrastructure as soon as possible. In the meanwhile, the 3GPP RAN3 is conducting research at present so that SON can be feasible to independently become a RAN logical entity or function. If SON is achieved in a logic entity form, it will be conducive to the uniform data acquisition and analysis of wireless side by SON, and achieving three core functions of SON on wireless side: parameter self-configuration, performance self-optimization and fault self-healing.

In addition to the NWDAF defined by 3GPP at the 5G core network side, the European Union 5G-MoNArch project group also suggests setting an independent AI analysis network function RAN-DAF on wireless side to conduct data analysis and decision of CU surface of 5G NR [99]. Due to the instantaneity of wireless side, such as scheduling management of wireless resources, etc., the intelligent analysis of wireless side needs real-time or quasi real-time decision. Therefore, the AI analysis based on real-time data shall be conducted in the local as far as possible to guarantee the real-time and dynamic performance optimization can be achieved. RAN-DAF will, as the AI and data analysis network function of radio access network, collect and monitor the data of UE and RAN on wireless site, including channel quality indicator (CQI), power level, path loss, radio link quality, radio resource usage, modulation and coding scheme (MCS), radio link control (RLC), buffer status information, etc. MoNArch suggests that RAN-DAF delivers those information to ran controller agent (RCA), which is an RAN side controller defined by MoNArch corresponding to the RIC defined by O-RAN. RAN-DAF and RCA jointly decide to optimize the wireless side quality, such as the flexible wireless resource control, selection of preferable radio access technology (RAT) of slice, management of cross-slice wireless resources, etc. Since RCA lacks the corresponding function in 3GPP, one relatively realistic selection is to only set RAN-DAF on RAN side which achieves the interconnection with core network side through cross-domain message bus in the form of SBA. The interface description of RAN-DAF, NWDAF and MDAF is shown in Figure 15. Currently, RAN-DAF of 5G-MoNArch hasn't been defined 3GPP standard yet, and it hasn't confirmed to undertake the control or management function of RAN or some SON functions in the future or not. It is suggested to take the independent logic virtual function of RAN-DAF or SON.

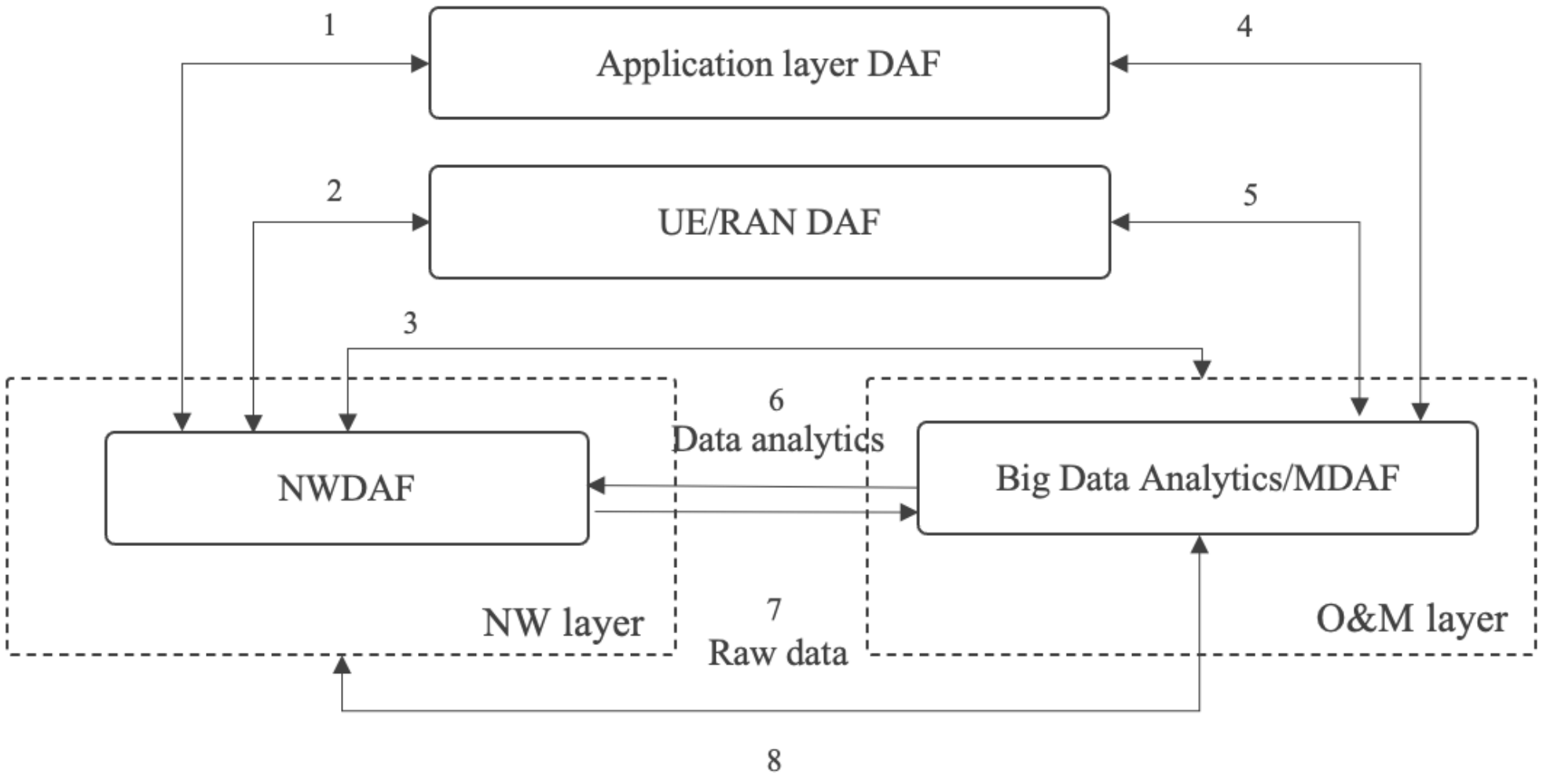

Figure 15 Data Analysis Framework Defined by 5G-MoNArch

RIC of O-RAN will continue to evolve and enhance, especially the intelligent strategy control for different App types, to help operators achieve the service orchestration based on App characteristics at the s service orchestration layer. RIC will perceive the App type, utilize the third-party xApps for the wireless resource management of the corresponding APP south bound according to the App features, and interact with the edge application server through north bound application program interface (API) according to the App type, as shown in Figure 16(a). RIC open function and enhanced wireless resources management function would include the data sharing among many O-RAN devices, SLA guarantee of wireless slice, wireless resource optimization of Internet of Vehicles / UAV, dynamic frequency spectrum sharing, combination with MEC to meet the business demands in vertical industry [130], as shown in Figure 16(b). service management and orchestration (SMO) defined by O-RAN, after its functional framework and interface gradually become mature, can support more powerful wireless cloud operation and maintenance management and non-real time RIC function.

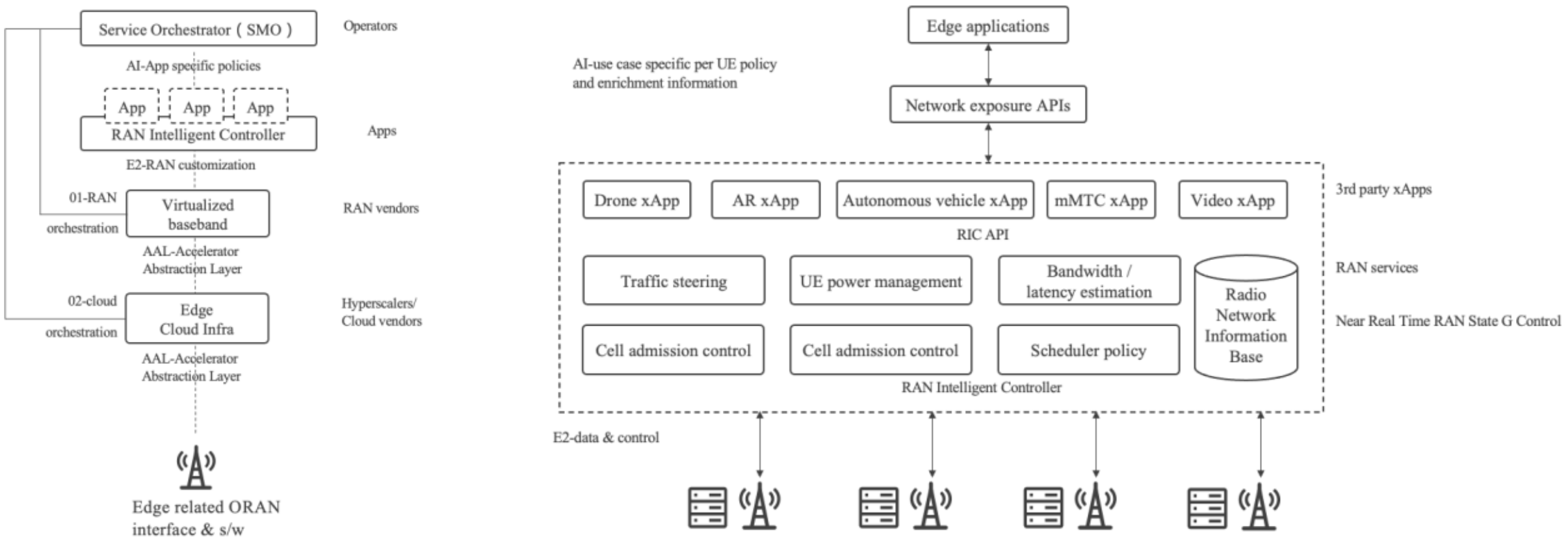

(a) Service orchestration based on O-RAN RIC (b) Capacity openness of O-RAN RIC

Figure 16 Service orchestration Based on O-RAN RIC and Capacity Opening of O-RAN RIC

- Core Network

NWDAF, as the AI network function of core network, will enhance the optimization of network performance and user experience in the future, and achieve the autonomous and intelligent service. With the comprehensive interconnection (such as other network function (NF), application function (AF), OAM, etc.) of network function interface around core network and the achievement of soft acquisition capacity of data, NWDAF will can comprehensively participate in the decision control of core network control plane in real time [131], for example, NWDAF cooperates with Network Slice Selection Function (NSSF) and policy control function (PCF), in which PCF can make strategy decision according to the slice level analysis results of NWDAF, and NSSF can make slice selection according to the load analysis of NWDAF.

One new case of NWDAF is "UE-driven analysis sharing". In this case, the information (such as user positioning information, Persona information, etc.) at UE terminal helps NWDAF make intelligent decision of network slice. Its key function is how NWDAF collects the UE-level information and how NWDAF conducts analysis with UE before providing analysis results for other NF. The relevant research projects of NWDAF also include the QoS safeguard based on NWDAF assistance, telephone traffic handling, personal mobility management, strategy decision, QoS adjustment, 5G edge computing, NF load balance, slice SLA guarantee, predictable network performance, etc. The achievement of those functions in NWDAF in the future will greatly improve the intelligentization of 5G core network. Cross-domain interaction of NWDAF is also worth boosting.

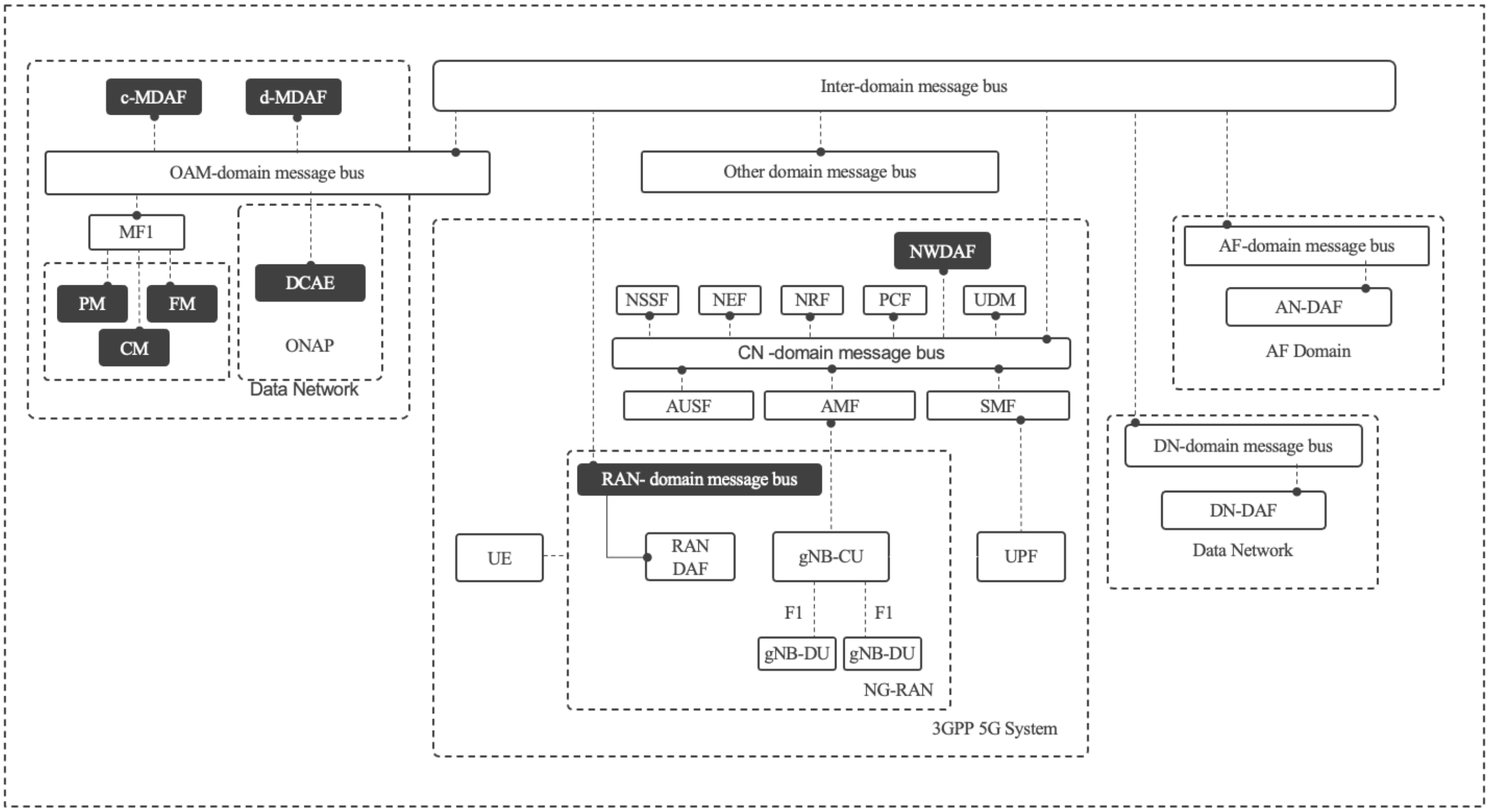

Figure 17 Forward looking of Telecommunications AI in Mobile Network Infrastructure

3GPP SA5 also researches how NWDAF empowers the analysis function to OAM or RAN. Besides, NWDAF will participate in the integration with MEC, and support the application in vertical industry through MEC to empower more applications in vertical industry. For the security problems concerned in the network field, it can be expected that NWDAF will also continue to enhance the relevant function, for example, NWDAF will monitor abnormal behavior of terminal or network, will promptly report to the relevant NF or OAM after finding abnormity, and take the corresponding protection strategy or measures. In respect of some key information transmission, such as AI algorithm model, NWDAF can combine blockchain and other technologies for traceability and security guarantee. Fusion architecture of NWDAF and MDAF/RAN DAF is shown in Figure 17.

- Transport Network

In the next decade, SDON/CON will combine AI more intimately, and will gradually achieve the "zero-touch" cognitive optical transport network and the fully automated management and control of optical network. The operational knowledge graph based on optical network will gradually become mature, through which the transmission problems can be located rapidly, transmission performance can be predicted and the transmission parameters can be optimized. The specific transmission indicators, such as modulation order, error correction, wavelength capacity, etc. can conduct the optimal configuration by utilizing AI technology to guarantee the transmission performance.

The IPv6-based application will gradually become mature, and the AI plays a key role in the SLA guarantee of network route and bearer network, and the deterministic network to achieve the intelligent IP network of IPv6 and even IPv6+ and to meet the personalized scenario demands of B5G/6G.

In respect of cloud-network integration, the computing power resources of cloud, network and edge will achieve completely distributed architecture, and provide seamless and high-quality computing power resources according to service requirements to offer resource guarantee to the high-level AI application at terminal and edge. After resilient computing force network/dynamic cloud-network integration became mature, new business model providing cloud computing service would occur, and the intelligent contract of block-chain, etc. can be utilized for security guarantee to solve the users' privacy problems and achieve the realizable capacities of network and computing resources.

Besides, network slice belongs to multi-domain technology across radio network, transport network, core network, etc. and each domain needs the coefficient cooperation, in which the transmission network is deemed as the physical basis to connect each domain, and the rational orchestration and support of transport resources are essential in the SLA guarantee of slice. It can be predicted that the AI will gradually become mature in the end-to-end slice SLA guarantee in the future.

- Terminal

The connection of terminal and network infrastructure conducts interaction through air interface and wireless network. Based on terminal, the AI in the face of future network infrastructure is mainly applied in the wireless perception function of terminal and chip, namely the AI technology based on terminal and chip perceives the wireless environment and contents to optimize the wireless access expenditure, time delay, etc. The achievement of wireless perception through terminal-based AI is embodied in the following 3 aspects [132]: frequency spectrum and access perception, namely certain terminal can detect other terminals' behavior so that 5G system performance can own better access and scheduling efficiency; content perception, namely the RF signal, sensor or telephone traffic behavior data are speculated and analyzed to obtain the users' contents, such as position, velocity, mobility used to optimize terminal performance and user experience; wireless environment perception, namely posture, action, certain object, etc. are detected through monitoring signal propagation and reflection mode, etc. to hasten new scenarios.

Terminal-based AI empowers the future 5G network system in the following 3 aspects: the first one is to enhance terminal experience, the intelligent beam forming and power consumption management can optimize velocity, robustness and battery's service life; the intelligent beam forming can be achieved through deep reinforcement learning, the position, velocity, other environments and application parameters are perceived to boost the robustness and velocity of network; In respect of the energy consumption saving, the performance and power consumption are weighed based on the AI contents by utilizing terminal. the second one is to improve 5G system performance, which is mainly reflected in intelligent link adaptation. The system velocity and frequency spectrum efficiency can be improved through location-based wireless interference prediction; Intelligent network load optimization can reduce the load that original data needs to transmit in the whole network through the terminal-based AI reasoning; For the intelligent seamless mobility, terminal-centered mobile management can predict the network handoff behavior and opportunity better through terminal AI and sensor. The third one is to improve the wireless security. The AI technology of terminal can be utilized to immediately detect and defend malicious base station fraud, malicious interference and other safety hazard behaviors in the local.

## 4.3 Forward Looking of Telecommunications AI in Network Management

Telecommunications AI will empower network management from many aspects, such as MDAF and ENI, Intent-based Networking or operator's operation and maintenance system defined in international standardization organizations. Besides, the network AI signaling system, network digital twins and orchestration system would also be developed greatly. The relevant prospect will be conducted in those aspects as below.

- MDAF

It can be predicted that 3GPP continues to evolve MDAF in SA5 working team from the standard perspective, and enhances the intelligent function of management plane related to operation and maintenance. In respect of the intelligence injection of SON, MDAF's application in many scenarios becomes mature gradually, such as coverage enhancement, resource optimization, fault detection, mobile management, energy saving, paging performance management, SON coordination, etc., for example, MDAF provides more accurate coverage analysis capacity and points out the causes of coverage problems, thus guiding the base station to adjust parameters and guaranteeing the users' service experience wouldn't be reduced. MDAF would also analyze the congestion condition of RAN User Plane more accurately, point out the congestion causes and provide the relevant strategy suggestions. MDAF would provide more accurate resource utilization-related analysis report, and offer the strategy suggestions to solve resource utilization problems. In respect of the key parameters of SLA, such as time delay, reliability, etc., MDAF would conduct accurate analysis and provide the suggestions about improvement of experimental performance. In respect of fault management, MDAF would conduct more accurate fault positioning and provide the relevant action suggestions. MDAF would also provide accurate strategy suggestions in the mobility management of users and improve the users' handover success rate and network efficiency. Stronger and more intelligent service is provided for slice management, which can finely manage various performance indicators of slice and guarantee the SLA parameters. Besides, the interaction of MDAF and network equipment (such as NWDAF) will also be improved.

- ETSI ENI

Currently, ENI system defines the functional architecture, but hasn't started the specific definition of interface. It can be predicted that the ENI working team will define the relevant interface so that the ENI has more reference significance in orchestration and application. Besides, based on the method for handling mechanism of data to combine ENI architecture, the orchestration method in operator system, the method for ENI to coordinate other intelligent network entities, the method to express and manage intention strategy, telemetry of flow information, etc., it can be predicted that the relevant work would also be carried out gradually, and the relevant scenario/function definition would become perfect gradually. The ENI of the next version will be enhanced accordingly in the intelligent application scenario and implementation.

- Intent-based Networking

Currently, China Telecom carries out the research on Intent-based Networking (IBN). 3GPP, ETSI, etc. start defining IBN [133][134], in order to define more intelligent network automation management mechanism. As described in 3GPP TR 28.812, the intent driven management (IDM) is sent by consumers to the provider (producer) of IDM, who conveys intention and then provides the relevant network configuration, as shown in Figure 18. In the meanwhile, the provider of IDM can monitor the network status and monitor whether intention of consumer is met, if not, the intention evaluation and parameter adjusted will be conducted again.

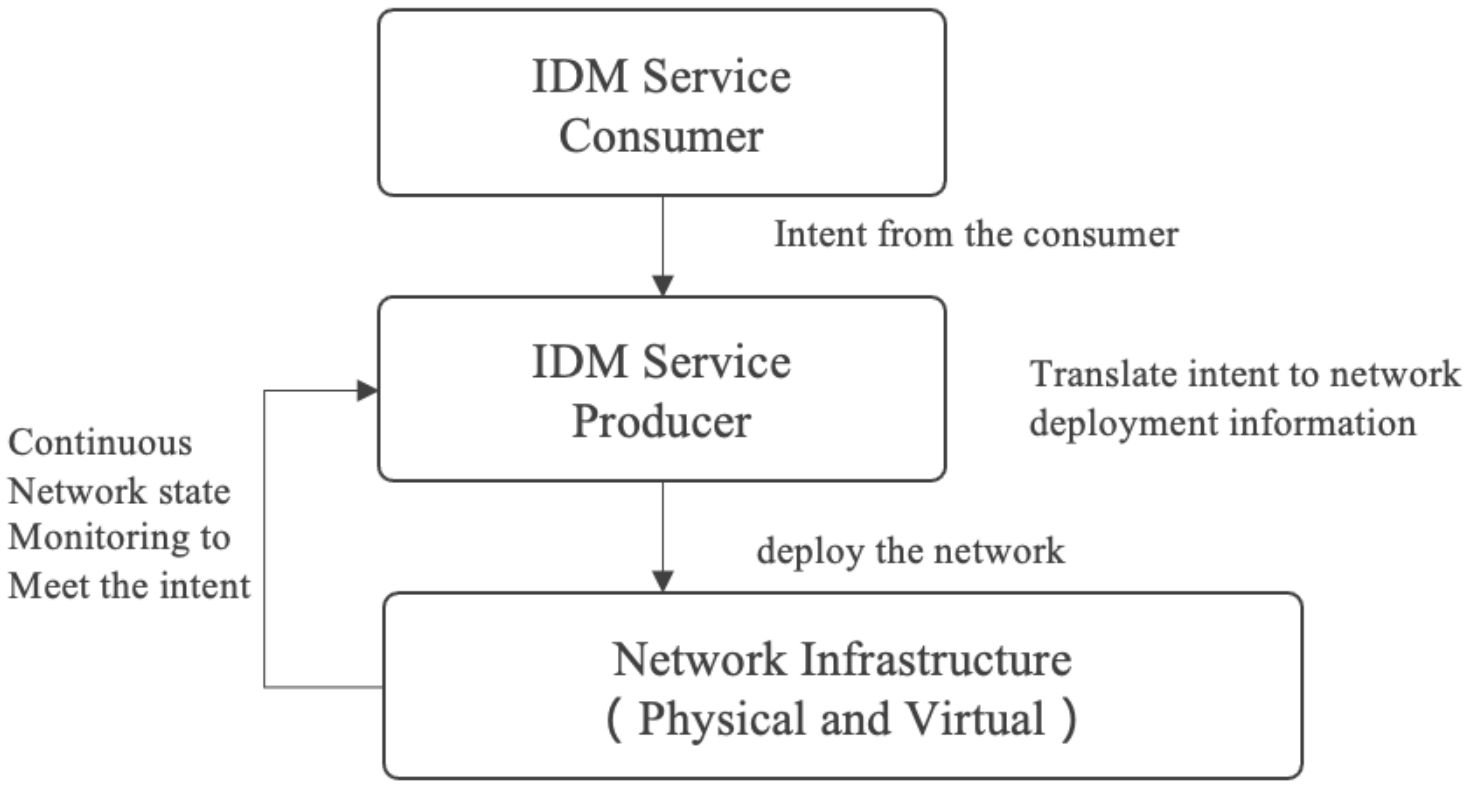

Figure 18 Intention Management

It can be predicted that the intent-driven management service will continue to evolve and be mature in the next decade, further reduce management complexity for operators and knowledge demands of underlying devices, and increase the network management efficiency in the scenario across many manufacturers. 3GPP SA5, ETSI ENI and other standardized organizations will continue to define rational and accurate intent expression, automatic mechanism and intent's life cycle management, etc. The application scenario would be enhanced and perfected gradually as expected in the network service opening, utilization and optimization of slice resource, slice performance guarantee, network optimization, network capacity management, network function deployment, etc.

- Network AI Signaling System

The network AI platform in 5G OSS middle office system can be considered as an AI platform and engine in the face of network management and operation function. Network AI platform needs to conduct interconnection & intercommunication with each south bound data acquisition network element or module and each north bound 5G OSS system (such as network orchestration, network performance, network resources and network fault) through a kind of standardized command system which is defined as network AI signaling system.

Different from the 4G/5G network signaling system which focuses on interconnection & intercommunication and interactive management of each network element, network AI signaling system is a whole set of standardized interconnection & intercommunication and AI interaction management command system between network AI middle office or platform and its each south bound and north bound interface and its each connecting network element, OAM module and each OSS system, including definition of network AI management command, uploading and distribution mechanism of network AI command among each interface, execution mechanism of network AI command stream, authentication and authorization mechanism of network AI, the training, reasoning, release and deployment mechanism of network AI algorithm, calculation of network AI algorithm, storage resource management, monitoring and safety of network AI command, etc.

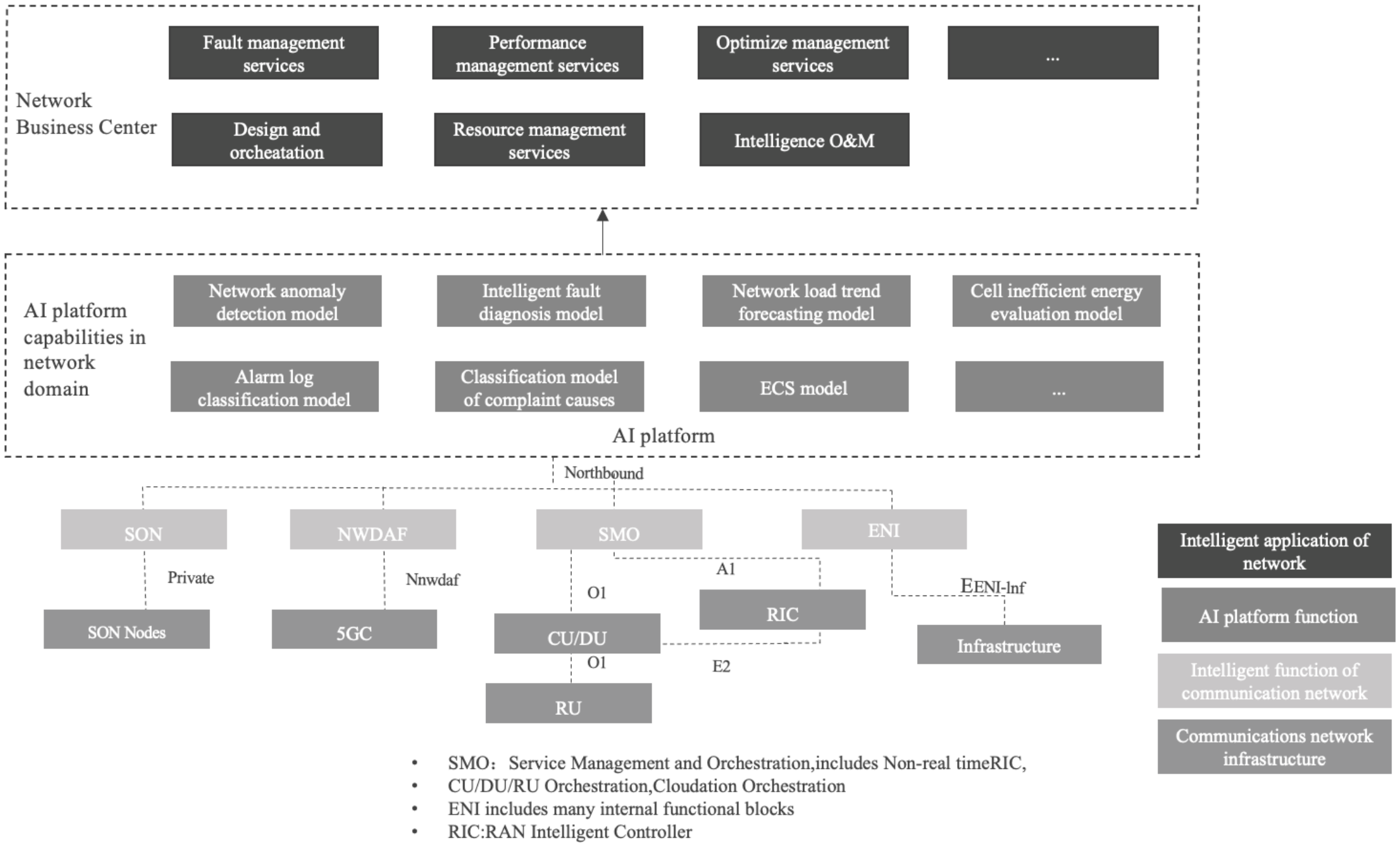

Figure 19 Network AI Signaling Interface

As shown in Figure 19, the typical network AI signaling interface includes the network AI middle office, AI data acquisition and AI command execution interface of 3GPP SON system, 3GPP NWDAF, O-RAN RIC, ETSI ENI and other systems, also includes each 5G OSS system of network AI middle office, such as network orchestration, network performance, network resources, network fault and other AI data acquisition and AI command execution interface. Figure 20shows a typical schematic diagram of network AI signal process for abnormal detection of cell telephone traffic and the corresponding network AI signaling packet structure.

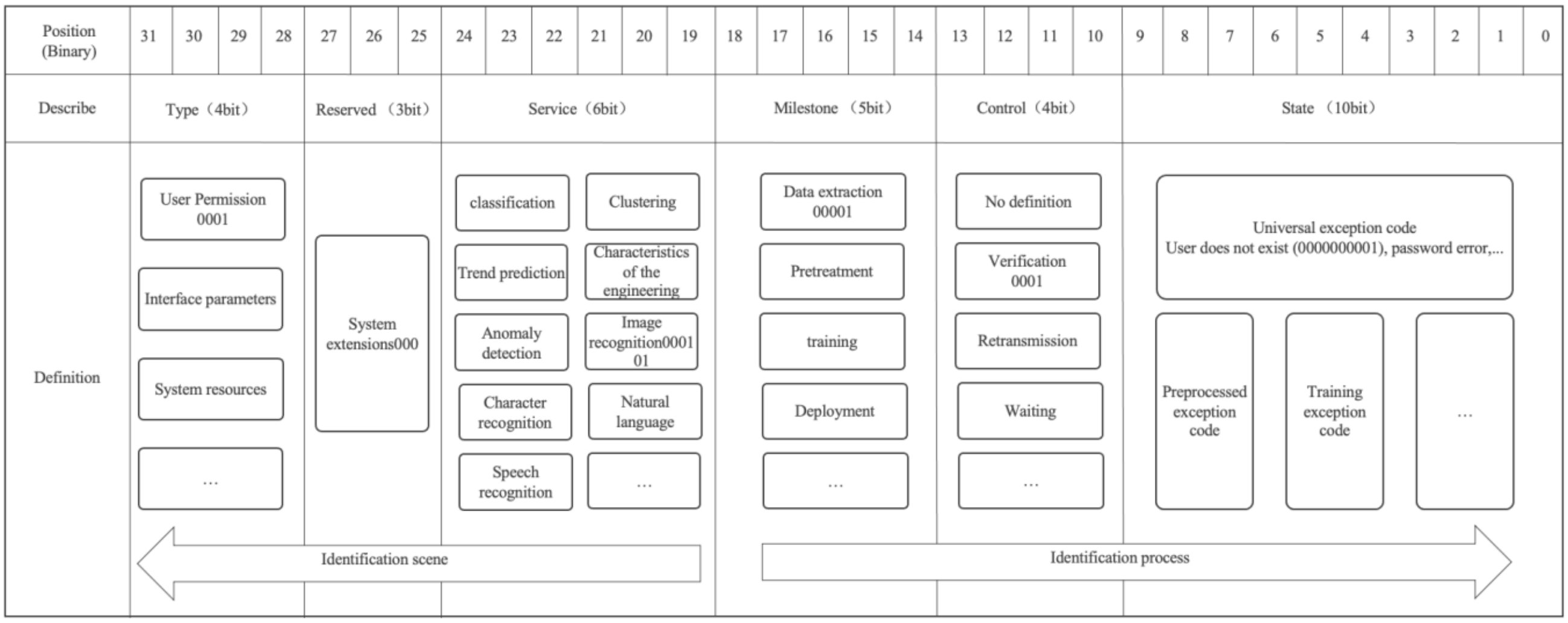

(a) AI Signaling Packet Structure.

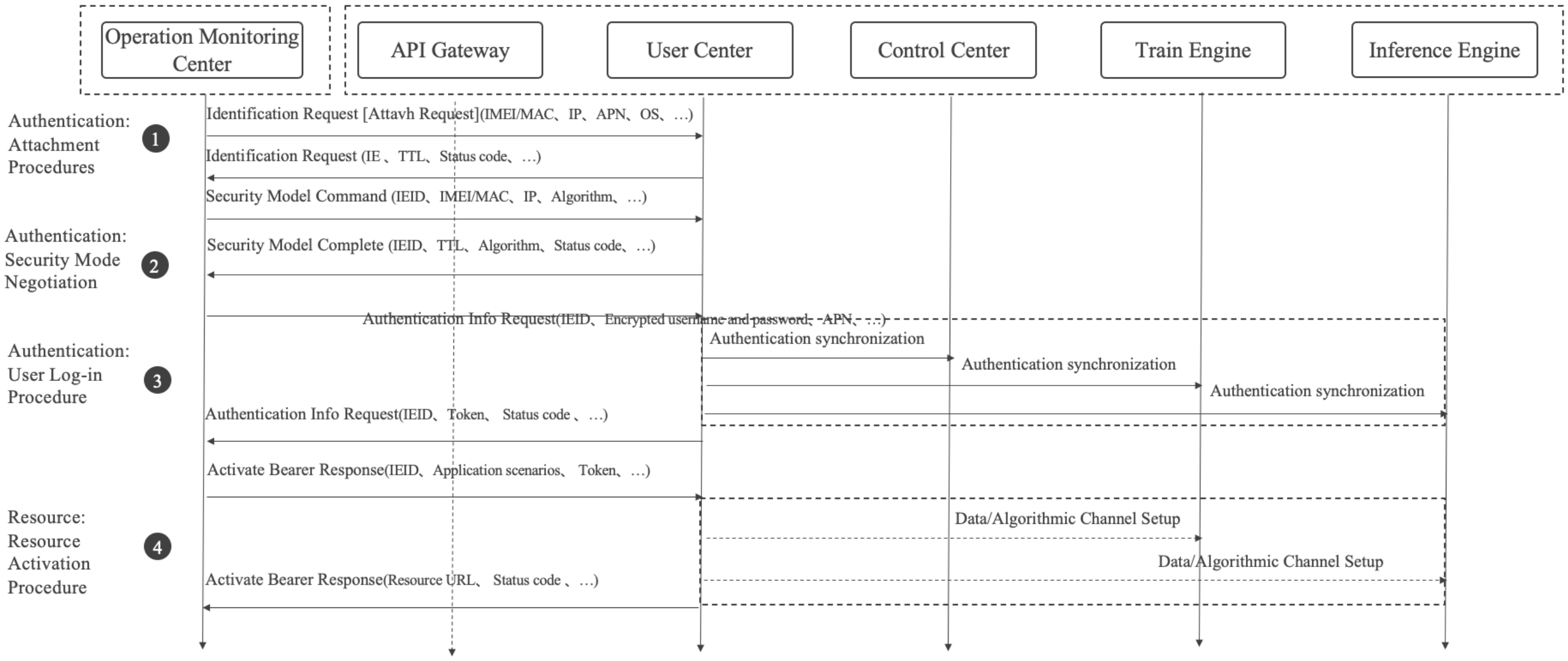

(b) Procedure of Network AI Signaling Process for Abnormal Detection of Cell Telephone Traffic

Figure 20 Typical network AI signaling packet structure and schematic diagram of network AI signal process for abnormal detection of cell telephone traffic

- Network Orchestration

Under the trend of network software definition and cloudification, network function (NF) management will be taken over by the management procedure defined by software, and will transform into virtualization management of shared computing and communication resource pool from the dedicated hardware. Therefore, for the traditional relatively solid OSS/BSS system, in addition to management function, its network orchestration capacity will play a crucial role in the future network technology evolution. ETSI preliminarily released the management and orchestration standard specification of network functions virtualization (NFV) in 2014, which defined the functional framework for MANO management and orchestration in NFV environment [135]. With the built-in of telecommunications AI in network orchestration, the industry is necessary to clearly describe the following evolution path of network orchestration in 5G and specify the logical and physical connection relation between AI function and network orchestration function. As shown in Section 3.2, network orchestration is a very important function in network management system of operators. The connection and construction of network, scheduling and orchestration of network resources, and network work process and business demand translation are corresponding to the network (topology) orchestration, network resource orchestration and network business orchestration, respectively. In 5G OSS, network service orchestration even can be independent of a sub-system and is mainly responsible for the orchestration and management of network slice service constituted by each virtualized network function (VNF) of 5G. The SLA-based slice intelligent management will be deemed as main application scenario of telecommunications AI in network service orchestration. The network service, such as guarantee and optimization of slice quality, also involves the instantaneity and intelligent scheduling support of upper service by underlying network resources. At this time, the network resource orchestration function plays a role in guaranteeing and supporting the upper network service orchestration function. The network resource orchestration shall achieve 3 elasticities in the future: the computing elasticity achieved in design and large-scale VNF; the orchestration-driven elasticity achieved through flexible setting of VNF; the slice perceived elasticity achieved through cross-slice resource supply mechanism [136]. Basic network resource orchestration function has been defined in ETSI MANO. ETSI ENI provides a reference architecture at present for the method to achieve the joint orchestration of network resources and network business based on SLA and ELA in the future and the method to specify the logical and physical connection of AI and network orchestration function, which can be continuously promoted and evolved in the industry [136], as shown in Figure 21.

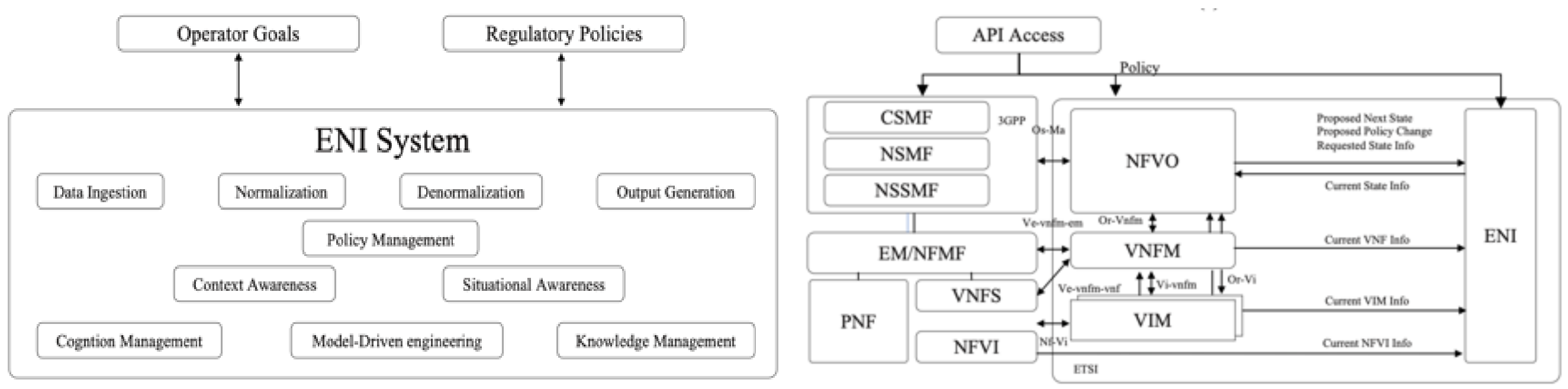

(a) ETSI ENI modular system architecture (b) Management and orchestration architecture

Figure 21 ETSI ENI Architecture and Its Interaction with ETSI NFV MANO Framework

In the meanwhile, before the existing traditional network with poor flexibility and weak automation capacity is replaced completely, how to coexist and coordinate with new network supporting network topology, resources and service intelligence orchestration, shall be considered by the industry, and the consistent action objectives shall be formed as soon as possible, including the integration of traditional network service (such as leased line) with automatic and relatively standardized orchestration rule process and new 5G network orchestration system.

At present, the network automatic and intelligent orchestration capacity of global operators is still at initial stage, and the technology and standard level shall be further perfected. It can be predicted that, with the deep integration of telecommunications AI and network orchestration system, the network (topology) orchestration, network resource orchestration and network service orchestration will be improved continuously. Based on the SLA or ELA optimization as convergence condition, the AI plays an important role in the translation and conversion and the proactive elastic matching between network service requirements and network topology construction and resource guarantee. In respect of standard level, it can be expected that 3GPP, ETSI and other standard organizations will also gradually improve the relevant scenarios, interfaces and process definitions.

### 4.4 Forward Looking of Telecommunications AI in Telecom Service

For the BSS intelligentization / financial management, it can be expected that AI will comprehensively empower customer management, package recommendation and financial intelligent management in the next decade, and achieve the transformation from the initial level to the advanced level. The AI will play a key role from the establishment of comprehensive customer experience with people-first to the establishment of business operation process with more efficiency for the telecommunication operators till the rapid innovation redemption of new business, new model and new technology. The customer operation, business operation, and business model and operation model will be described as below.

- Customer Operation

From the perspective of customer operation, telecommunication operators have arrived the comprehensive customer experience phase from customer-centric to customer experience-centric. The telecommunication operators not only pay attention to provide customers with the refined operation and service satisfaction improvement in marketing, sales, and service process, but also add the customers experience in network and business usage to total experience system as important indicators for uniform evaluation. The customers' experience in marketing, network service and business use is collected, converged, associated and mined to establish the customer-centric total experience. On that basis, the AI algorithm assistance is combined to further meet the customer demands in more refined scenarios, and the intelligent and automatic interaction capacity is further improved to form the customer-centered comprehensive customer experience.

- Business Operation

From the perspective of business operation, the telecommunication operator has basically accomplished end-to-end digitalization upgrade of the whole process, and is implanting big data, AI, etc. into the existing process to accomplish the intelligentization of business processing process and further improve business operation efficiency. With the introduction of RPA, intelligent business process management suites (IBPMS), etc., it is predicted that the business process with manual intervention will be further decreased, process operation efficiency will be further improved, and the cost will be further reduced in the future. As AI is introduced to risk control system of the operator, the income guarantee capacity is further improved, and the arrear risk will be further reduced. The operators can combine own predicted risk control to carry out more innovative businesses, which will further promote more healthy and upward cash flow of telecommunication operator.

- Business Mode and Operation Mode

From the perspective of business mode and operation mode of current telecommunication operator, it can't meet the high-speed development requirements using heavy asset input and the volume-based billing method for the mass market. The traditional manual method which adopts large granularity to subdivide customer group for product and pricing on the mass market is very difficult to meet the scenario-based and personalized requirements of customers in the future. Therefore, operators shall take full advantage of AI capacity for market segmentation with finer force and business operation in virtual expert/personal assistant method, coordinate intelligent workflow and risk control system, promote business products and services of "one customer with one strategy" and "everyone with own opinion" to customers, and further improve innovation through business product innovation and operation model innovation.

## 4.5 Forward Looking of Telecommunications AI in Cross-domain Intelligentization

For the integration intelligentization cross B&O domains, it is predicted that the architecture and function of CEM and PCF+ will have continuous development in the next decade, and the customer experience perception system will evolve from SLA to ELA.

- Enhancement of Customer Experience Perception and Evolution from SLA to ELA

Since the 1990s, the QoS [137][138] defined by international standard organizations such as ITU, ISO, etc. was adopted by most telecom operators and used as the SLA signed with customers. Traditional QoS is driven by technology, especially the network and service performance indicator used to define service quality [139][140], namely the indicators in network is included in the SLA. Since traditional QoS can't truly reflect the users' experience and feeling in use of network service, the service quality system has gradually developed and evolved into the user-centric QoE in recent years. ITU defined QoE as the acceptable overall indicator of certain application or service which can be subjectively perceived by end users [140].

In the current SLA system, telecom operators provide users with the “best-effort” service, for example, the bandwidth 100 Mbit/s often means "the maximum velocity reaches 100 Mbit/s in ideal condition". Such best effort service and KPIs from the technical perspective are very difficult to directly associate the users' experience quality. In the meanwhile, when users pay for services, they often subjectively affirm the stability and quality of service shall be more than or equal to their expected costs. With CEM QoE algorithms becomes gradually mature in the future, telecom operator will utilize QoE algorithms for predictive evaluation and proactive management of users' experience expectation. Therefore, it is imperative for traditional customer-oriented SLA system to evolve into the QoE-based ELA. ELA is defined as "a kind of special SLA, namely a kind of quality grade-based consensus formed by users in use and experience of certain service when customers clearly know certain service" [132].

Figure 22 describes the correspondence of service quality guarantee between the communication operator, content provider, toB customer and the toC customer [132]. ELA aims at toC customers and operators and content providers. In current business mode, ELA mainly refers to the network service and the mixed ELA mode based on content quality, namely a whole set of ELA system contains the network service quality and the experience quality on application layer. Currently, customers are very difficult to clearly define the experience reduction at network side or application side. While in the future, the ELA system shall be based on Figure 22 (b), namely construction of independent ELA system for toC customer in network side and application side. For the users, breaking all experience quality black box attributing to "not good network" in QoE perception area can make the advantages and disadvantages of experience quality provided by different service providers. In the meanwhile, content providers can also clearly define various QoS requirements in the ISP SLA of communication operator in order to guarantee better QoE for toC customers.

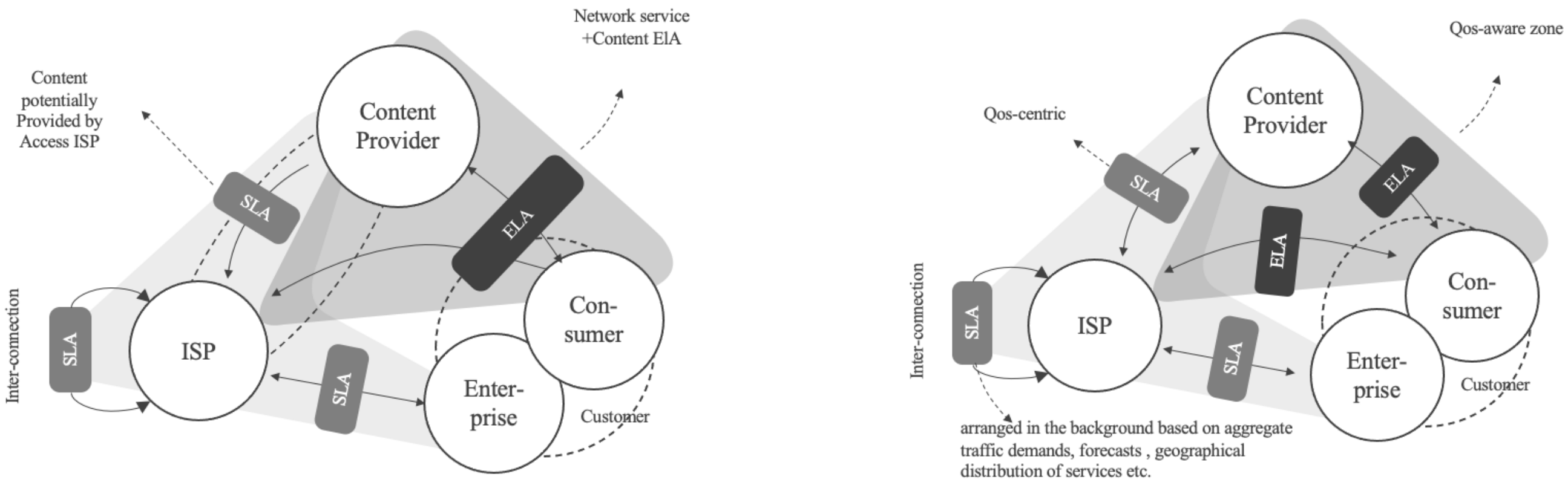

(a)Specific service contracts　　　( b ) Service-independent agreements

Figure 22 Construction of ELA Ecology through SLA and QoE Concept

- CEM Architecture Evolution

Since CEM data are from many domains: OSS, BSS and application layer, "data compliance" and "data silo" make the data aggregation of CEM in many aspects very challenging [141]. In order to balance the demands of "privacy and security" and "data fusion", trans-regional construction of CEM model is proposed. The trans-regional modeling is achieved in two most common methods: transfer learning (TL) and federated learning (FL). A pre-trained model (PM) with strong generality is obtained through large-scale training of massive data, then the pre-trained model is moved to small-scale data application scenario, and small-scale data is utilized for fine tuning of pre-trained model, thus achieving the significant improvement of model performance [141]. Such method can be used to construct the CEM model, as shown in Figure 23(a), the basic PM of CEM is obtained through the data model training of certain telecom operator, then the scenario service data are utilized for fine tuning of PM to obtain the scenario-based CEM, which avoids the direct data aggregation.

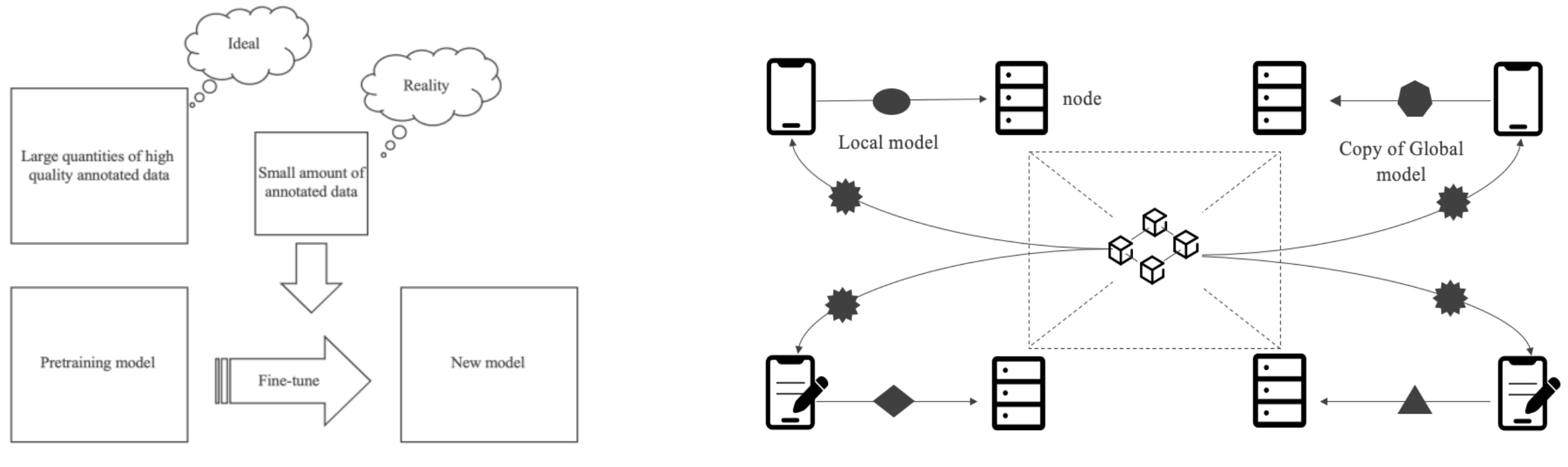

(a) CEM modeling based on Transfer Learning [141]　　　(b) CEM modeling based on Federated Learning [22]

Figure 23 CEM modeling methods based on Transfer Learning and Federated Learning

In 2017, a kind of joint modeling method for fragmentation data- Federated Learning was proposed [22], Federated Learning is very suitable for the CEM modeling scenario. Federated Learning adopts the modeling thought of "data keeps unchanged while model changes" to avoid the privacy and security of customer experience data very well. In the meanwhile, it is also a kind of multiparty secure computing technology to solve the data silo problem very well, as shown in Figure 23(b). In the joint modeling process, the data of all parties are kept in the local. Therefore, the customer experience perception method based on Federated Learning technology is safe, controllable and stable, and greatly improves the accuracy of joint model and the comprehensiveness of evaluation.

- Evolution of PCF+

Based on the subsequent evolution version of PCF+, it can be predicted that PCF+ provides users with more accurate, real-time and differentiated strategy control through the interaction with OSS domain and BSS domain, as shown in Figure 13 PCF+ can unify BSS domain strategy, and guarantee the reliability and instantaneity of service in the process of providing strategy control. For the strategy control of network side, PCF+ can control the user QoS parameters dynamically in real time and conduct the session management to guarantee the service quality. Since PCF+ has the capacity to pull through O/B domain, it becomes more personalized in strategy control of users and can further explore the scenarios with added value to achieve the business vision of B5G/6G.

The PCF+ can be deployed in OSS domain or BSS domain to serve different domains. The PCF+ among different domains can conduct mutual interaction to obtain the data information in other domains. PCF+ can also be hierarchical deployed according to actual condition of network.

### 4.6 Forward Looking of Telecommunications AI in Private Network Application

One of core values of 5G is the dedicated application for to-B enterprise. It can be predicted that, in the next decade, the telecommunications AI will help enterprises achieve advance intelligent even completely intelligent private network function in vertical industry applications, such as Internet of Vehicles, Intelligent Manufacturing, High-definition Video /VR/AR, Telemedicine and Smart City.

Some use cases about 5G private network in the current industrial Internet effectively promote the application and ecological construction of 5G in vertical industry filed, and private network with different deployment forms can also be applicable for diversified network demands, but there is still great gap away from the demands in vertical industry. It can be predicted that, in the next decade, telecommunications AI can completely meet the requirements of vertical industry in high-quality communication and network safety through combination with MEC and business, and AI algorithms, such as Federated Learning, Transfer Learning, etc., will help private network solve data privacy and security, data volume deficiency, etc., and AI technology is utilized to perceive service change, optimize wireless network parameters, thus guaranteeing the service transmission quality, and truly changing current private network into high-performance, safe and reliable private network.

# 5 EXPECTABLE FUTURE: COMPREHENSIVE PROMOTION OF COMMUNICATION INTELLIGENTIZATION IN THE NEXT DECADE

In the evolution process of 6G, 3GPP will abide by the rhythm for one Release every 2~3 years, and will be expected to evolve to R21 from R17 in the next decade. Currently, ITU-R also starts researching the 6G technology trends and vision, and is hopeful to specify formal standard of 6G in 2027-2028. In the evolution process of 3GPP and ITU-R technical standards, every function domain (core network, wireless, transport, network management, business support, network application, etc.) of mobile communication will gradually reach the intelligent advanced level of communication ecosystem in the next decade through different levels of coupling development with AI. As shown in Figure 24 it will achieve the communication autonomy and fully intelligentization objective in B5G/6G vision in the end.

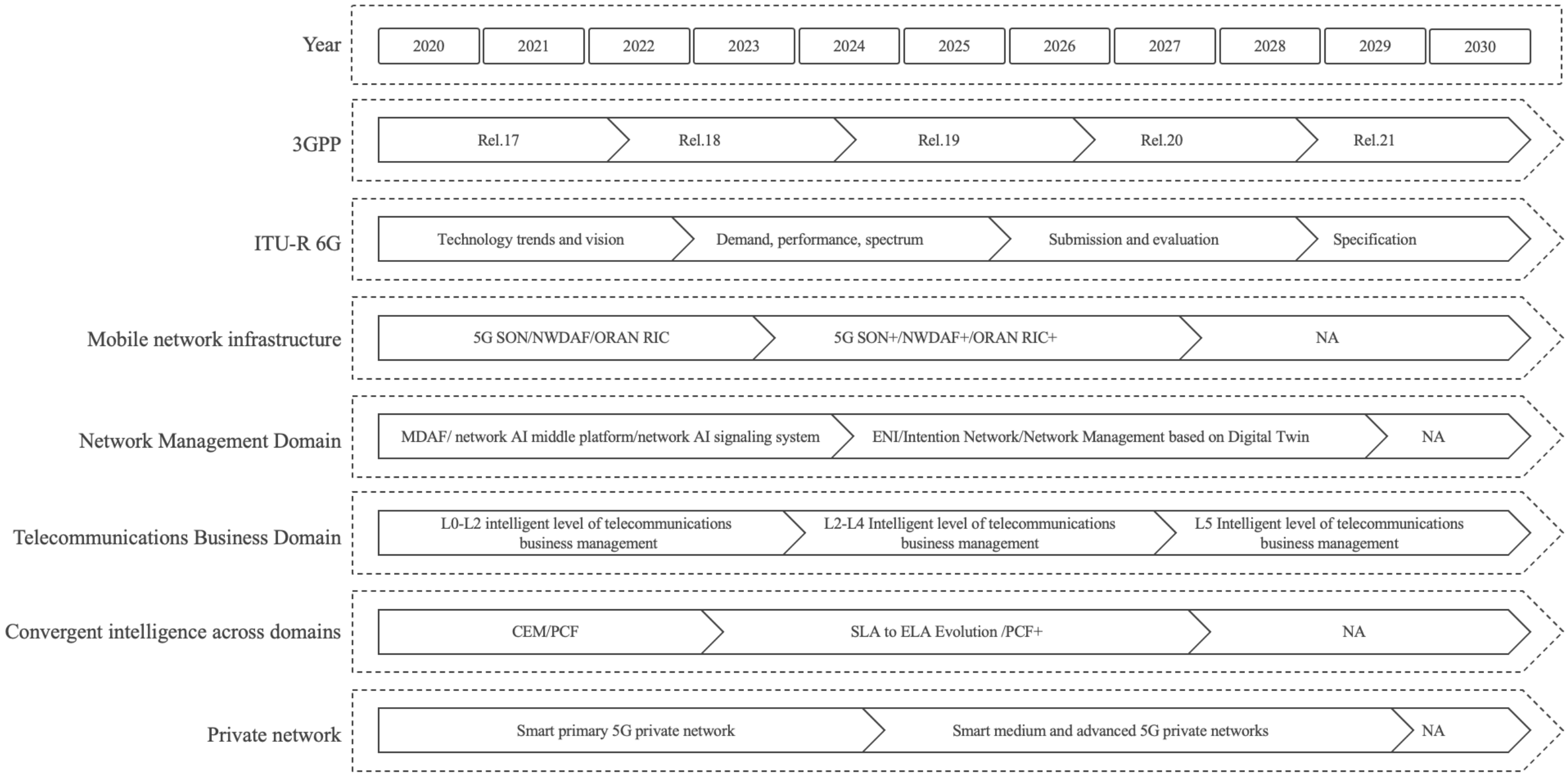

Figure 24 Forward Looking of Telecommunications AI in the Next Decade

AI aims at the communication network infrastructure. In the future 3~5 years, NWDAF will be gradually mature for commercial use in 5G core network; Radio and core network optimization will also achieve the AI-driven intelligent network optimization objective by virtue of SON. The commercial deployment method of SON is likely to adopt independent SON system for orchestration or integration in 5G OSS system, and the definition of RAN-DAF in independent network element form hasn't been determined. In the future 5~10 years, with the gradual commercial use of O-RAN, RIC, as intelligent controller with open radio network, will also achieve the commercial deployment.

AI aims at network management and operation. In the future 3~4 years, MDAF achieve some data analysis functions for network management. With the construction of network middle office system, the network AI middle office will achieve commercial deployment in 5G OSS system of some operators. Network AI signaling system, as the interaction language of AI and network, will inject the AI capacity into network ecosystem. In the future 5~10 years, intent-based network and experience and perception system of ETSI ENI will be applied in the network infrastructure of beyond 5G (B5G). Digital twins technology will combine with network simulation and AI to achieve the intelligent management of network planning, network construction, and network optimization & maintenance in the whole life cycle of network.

AI aims at the telecommunication business and support, and technology middle office system constructed by some telecom operators will achieve comprehensive commercial use and become mature in the future 3~5 years. AI platform, as the intelligence injection and empowerment of AI in BSS domain, will comprehensively promote the intelligentization of customer operation and business operation. Since there is similar common application experience in vertical industry, some dedicated fields (such as intelligent customer service, intelligent marketing, intelligent recommendation, etc.) involved in telecom business will accelerate development in the future 5~10 years and may achieve the advance intelligent level at L4 or L5.

AI aims at the intelligent integration across BSS and OSS domain. CEM and PCF will develop along the evolution path with integration of BSS and OSS domain, in which CEM will achieve the closed-loop management of customers' network and service experience perception in the whole life cycle journey of customers in combination with the network and service data. Since the cross-domain data use is involved, future architecture of CEM can be achieved through Federated Learning. Customer experience and perception management system would evolve from SLA to ELA system. PCF can provide accurate, real-time and personalized strategy and service for the network, service and customers through interaction with OSS domain and BSS domain.

The private network in vertical industry will be at the initial stage of commercial construction in the future 3~4 years, and its main deployment mode is achieved in the form of virtual private network. Therefore, the application of AI in virtual private network will focus on the SLA guarantee of 5G private network slice, intelligent scheduling and optimization of slice resources, coverage and performance optimization of wireless private network, etc. In the future 5~10 years, mixed private network and independent private network will be gradually deployed and mature. The AI application in independent or mixed private network will focus on the accurate QoS guarantee of toB business, real-time evaluation optimization of toC business perception experience, intelligent network AIOps, etc. Besides, AI technology conducts the adaptive adjustment of parameter setting of application layer through interaction with private network application platform multi-access edge platform (MEP) in vertical industry to guarantee the service quality of edge application. AI mainly focuses on the intelligent performance, quality and operation and maintenance guarantee at the primary intelligent stage of industry private network, and will focus on the intelligent control and management with high reliability, low latency and massive connections at the advanced intelligent stage.

In the future, the telecommunications AI system will further cement the security, robustness, interpretability, etc. Especially the technical combination of Federated Learning, block-chain and privacy computing will develop in each communication ecosystem as expected to solve the data silo and security and privacy problems between communication ecosystem and vertical industry: The Federated Learning model involves the training of multi-party data, and the Federal Center is responsible for the secret key management and model management by regularly auditing the Federal Center; Block-chain technology solves the consensus and credibility problem; model training, model inference and data ID alignment have use block-chain and the trustworthy network with multi-party cooperation is achieved through smart contract, consensus calculation, etc., and the block can replace the Federal Center in the multi-party federal computing network.

In the face of 6G, telecommunications AI will promote the multi-dimensional ecosystem intelligent integration of Space-air-ground-ocean. Since the expansion of 6G communication ecosystem in spatial dimension will generate more scenarios, the telecommunications AI will solve the NP-hard (non-deterministic polynomial-time hard) problems of many sub ecosystems with polynomial complexity such as cross-layer optimization, joint optimization, etc.

Currently, architectural integration and functional application of AI in communication ecosystem has achieved standard definition in 3GPP, ITU-R and ETSI. The commercial process of telecommunications AI in 5G is at the early stage, especially telecommunications AI-related network function (such as 3GPP NWDAF or O-RAN RIC) is at the test stage and is rarely used for commercial purposes in 5G network. At present, telecommunications AI is widely applied in network management and business management & application, and obtains periodic achievements. The authors appeal to the telecom industry builders to make more openness in network standardized function, give AI the opportunity for comprehensive empowerment and intelligence injection for network infrastructure, OSS and BSS etc., and comprehensively release the 5G and AI as GPT portfolio potential in the telecom ecosystem and vertical application.